\documentstyle[epsf,epsfig]{mn}
\newif\ifAMStwofonts
\newcommand{\kms} {km s$^{-1}$\ }
\ifoldfss

  \newcommand{\rmn}[1] {{\rm #1}}

  \ifCUPmtlplainloaded \else
    \NewTextAlphabet{textbfit} {cmbxti10} {}
    \NewTextAlphabet{textbfss} {cmssbx10} {}
    \NewMathAlphabet{mathbfit} {cmbxti10} {} % for math mode
    \NewMathAlphabet{mathbfss} {cmssbx10} {} %  "   "    "
  \fi
  \ifAMStwofonts
    \ifCUPmtlplainloaded \else
      \NewSymbolFont{upmath} {eurm10}
      \NewSymbolFont{AMSa} {msam10}
      \NewMathSymbol{\upi}     {0}{upmath}{19}
      \NewMathSymbol{\umu}     {0}{upmath}{16}
      \NewMathSymbol{\upartial}{0}{upmath}{40}
      \NewMathSymbol{\leqslant}{3}{AMSa}{36}
      \NewMathSymbol{\geqslant}{3}{AMSa}{3E}

       \let\le=\leqslant
       \let\ge=\geqslant
    \fi
  \fi
\fi % End of OFSS

\ifnfssone
  \newmathalphabet{\mathit}
  \addtoversion{normal}{\mathit}{cmr}{m}{it}
  \addtoversion{bold}{\mathit}{cmr}{bx}{it}
  \newcommand{\rmn}[1] {\mathrm{#1}}

  \newmathalphabet{\mathbfit} % math mode version of \textbfit{..}
  \addtoversion{normal}{\mathbfit}{cmr}{bx}{it}
  \addtoversion{bold}{\mathbfit}{cmr}{bx}{it}
  \newmathalphabet{\mathbfss} % math mode version of \textbfss{..}
  \addtoversion{normal}{\mathbfss}{cmss}{bx}{n}
  \addtoversion{bold}{\mathbfss}{cmss}{bx}{n}
  \ifAMStwofonts
    \ifCUPmtlplainloaded \else
      \UseAMStwoboldmath
      \makeatletter
      \new@mathgroup\upmath@group
      \define@mathgroup\mv@normal\upmath@group{eur}{m}{n}
      \define@mathgroup\mv@bold\upmath@group{eur}{b}{n}
      \edef\UPM{\hexnumber\upmath@group}
      \new@mathgroup\amsa@group
      \define@mathgroup\mv@normal\amsa@group{msa}{m}{n}
      \define@mathgroup\mv@bold\amsa@group{msa}{m}{n}
      \edef\AMSa{\hexnumber\amsa@group}
      \makeatother
      \mathchardef\upi="0\UPM19
      \mathchardef\umu="0\UPM16
      \mathchardef\upartial="0\UPM40
      \mathchardef\leqslant="3\AMSa36
      \mathchardef\geqslant="3\AMSa3E

       \let\le=\leqslant
       \let\ge=\geqslant
    \fi
  \fi
\fi % End of NFSS release 1

\ifnfsstwo
  \newcommand{\rmn}[1] {\mathrm{#1}}

  \DeclareMathAlphabet{\mathbfit}{OT1}{cmr}{bx}{it}
  \SetMathAlphabet\mathbfit{bold}{OT1}{cmr}{bx}{it}
  \DeclareMathAlphabet{\mathbfss}{OT1}{cmss}{bx}{n}
  \SetMathAlphabet\mathbfss{bold}{OT1}{cmss}{bx}{n}
  \ifAMStwofonts
    \ifCUPmtlplainloaded \else
      \DeclareSymbolFont{UPM}{U}{eur}{m}{n}
      \SetSymbolFont{UPM}{bold}{U}{eur}{b}{n}
      \DeclareSymbolFont{AMSa}{U}{msa}{m}{n}
      \DeclareMathSymbol{\upi}{0}{UPM}{"19}
      \DeclareMathSymbol{\umu}{0}{UPM}{"16}
      \DeclareMathSymbol{\upartial}{0}{UPM}{"40}
      \DeclareMathSymbol{\leqslant}{3}{AMSa}{"36}
      \DeclareMathSymbol{\geqslant}{3}{AMSa}{"3E}

       \let\le=\leqslant
       \let\ge=\geqslant
    \fi
  \fi
\fi % End of NFSS release 2

\ifCUPmtlplainloaded \else
  \ifAMStwofonts \else % If no AMS fonts
    \def\upi{\pi}
    \def\umu{\mu}
    \def\upartial{\partial}
  \fi
\fi

\title[Massive stars in M31]
{Chemical abundances and winds of massive stars in M31: a B-type supergiant 
and a WC star in OB\,10}
\author[Smartt et al.]
       {S.J. Smartt,$^1$ P.A. Crowther,$^2$ P.L. Dufton,$^3$\thanks{
       On leave of absence at the Isaac Newton Group
       of Telescopes, Apartado de Correos 368, E-38700, 
       Santa Cruz de La Palma} D.J. Lennon$^4$, R.P. Kudritzki$^{5,8}$, 
 \newauthor  A. Herrero$^{6,9}$, J.K. McCarthy$^7$, F. Bresolin$^{5}$. 
       \\
       $^1$Institute of Astronomy, University of Cambridge, Madingley Road,
	Cambridge CB3 0HA 
	\\
	$^2$Department of Physics and Astronomy, University College London,
	Gower Street, London WC1E 6BT
	\\
	$^3$The Department of Pure and Applied Physics, The Queen's University
	of Belfast, Belfast BT7 1NN
        \\
        $^4$ The Isaac Newton Group of Telescopes, Apartado de Correos 368, 
        E-38700, Santa Cruz de La Palma, Canary Islands, Spain
        \\
        $^5$ Institut f\"{u}r Astronomie und Astrophysik der Universit\"{a}t
        M\"{u}nchen, Scheinerstr.1, D-81679, M\"{u}nchen, Germany
        \\
        $^6$ Instituto de Astrof\'{i}sica de Canarias, E-38200, La Laguna, 
        Spain
        \\
        $^7$ Pixel Vision Inc., Advanced Imaging Sensors Division 4952 
	Warner Avenue, Suite 300, Huntington Beach, CA 92649
        \\
        $^8$ Institute for Astronomy, University of Hawaii at Manoa,
	2680 Woodlawn Drive, Honolulu, Hawaii 96822
        \\
        $^9$ Departamento de Astrofisica, Unversidad de La Laguna, E-38071 La
        Laguna, Tenerife, Spain }
        
\date{Accepted 
      Received ;
      in original form }

\pagerange{\pageref{firstpage}--\pageref{lastpage}}
\pubyear{2000}

\begin{document}

\maketitle

\label{firstpage}

\begin{abstract}
%pac
We present high quality spectroscopic data for two 
massive stars in the OB\,10 association of M31, OB\,10--64 (B0\,Ia) 
and OB\,10-WR1 (WC6). Medium resolution spectra of both stars 
were obtained using the  ISIS spectrograph on 
the William Hershel Telescope. This is supplemented with HST-STIS UV
spectroscopy and Keck~I HIRES data for OB\,10--64. 
A non-LTE model atmosphere and abundance analysis for OB\,10--64 is presented 
indicating that this star has
similar photospheric  CNO, Mg and Si abundances as solar neighbourhood 
massive stars. A wind analysis of this early B-type supergiant
reveals a mass-loss rate of $\dot{M}=1.6\times10^{-6}M_{\odot} 
{\rm yr}^{-1}$, and $v_{\infty}= 1650$ km\,s$^{-1}$. 
The corresponding wind momentum is in good agreement with the
wind momentum -- luminosity relationship found for Galactic early B
supergiants.

Observations of OB\,10-WR1 are analysed using a non-LTE, line-blanketed
code, to reveal approximate stellar parameters of
log$L/L_{\odot}\sim$5.7, $T_{\ast}\sim$75 kK, 
$v_{\infty}\sim 3000$ km\,s$^{-1}$, $\dot{M}/(M_{\odot} 
{\rm yr}^{-1}) \sim10^{-4.3}$ adopting a clumped wind with a filling 
factor of 10\%. Quantitative comparisons are made with the Galactic 
WC6 star HD\,92809 (WR23) revealing that OB\,10-WR1 is 0.4 dex more
luminous, though it has a much lower C/He ratio ($\sim$0.1 versus 0.3 for 
HD\,92809). Our study represents the first detailed, chemical model 
atmosphere analysis for either a B-type supergiant or a WR star in 
Andromeda, and shows the potential of how such studies can provide 
new information on the chemical evolution of galaxies and the 
evolution of massive stars in the local Universe. 
\end{abstract}

\begin{keywords}
stars:abundances -- stars: early-type -- stars: Wolf-Rayet -- galaxies: M31
 -- stars:winds
\end{keywords}

\section{Introduction}

%pac
The observation and analysis of 
hot, luminous stars in the Milky Way and other Local Group galaxies
allows the investigation 
of stellar evolution and mass-loss within different environments. 
For example, Wolf-Rayet (WR) stars, the
chemically evolved descendents of massive 
OB stars, provide keys to our understanding
of massive stellar evolution and nucleosynthetic processes, e.g. Dessart
et al. \shortcite{Des00}. 

Studies of the 
physical nature of the A and B-type supergiants can also place
constraints on evolution models in the upper main-sequence regions,
e.g. Venn \shortcite{venn95}, McErlean, Lennon \& Dufton \shortcite{McE99}. 
A further use of the A and B-type supergiants is in using 
photospheric analysis to constrain the chemical composition of 
the interstellar medium in their host galaxies. They are visualy 
the brightest quasi-stable objects in galaxies,
and being descendants of massive OB-type main-sequence objects
have small lifetimes (5--20\,Myr). 
Many elements visible in their photospheres should not be
affected by internal mixing and contamination with 
core processed material (e.g. O (only slightly affected in
AB-types), Mg, Si, S, Al, Ca, Fe, Ti, Cr, Ni; 
Venn et al. 2000, Smartt, Dufton \& Lennon 1997). Further they 
can also provide new information on the distances of galaxies through the 
detailed study of the strengths of their radiatively driven winds using
the wind momentum -- luminosity relationship (WLR, e.g. Kudritzki et al. 
1999). An investigation of the stellar wind properties of O-,
B- and A-supergiants in Local Group galaxies with well defined distances
will, therefore, allow us to test the concept of the WLR and its applicability
for distance determinations.

Numerous surveys have identified OB and WR stars beyond the Magellanic Clouds,
e.g. Massey et al. \shortcite{Mas86}; Moffat \& Shara \shortcite{MS87}, 
although little quantitative analysis has been carried out to date. 
The only detailed studies of Wolf-Rayet stars beyond the Magellanic Clouds
have been studies of  late WN stars in M33 by Smith et al. 
\shortcite{SCW95} and Crowther et al. \shortcite{Cro97}.
Bianchi et al. \shortcite{Bia94} have published UV 
spectroscopy of OB-type supergiants in M31 and M33, while Monteverde et 
al. \shortcite{Mon97,Mon2000} derived the oxygen abundance gradient
in M33 from studies of B-type supergiants. An indication of the capabilities 
of  8m class telescopes has been presented by Muschielok et al. 
\shortcite{Mus99} for three B-type supergiants in NGC\,6822 using VLT.
Recently, Venn et al. \shortcite{venn2000}  
have analysed four A-type supergiants in M31, 
the first detailed stellar analysis in this galaxy.

The present work demonstrates the capabilities of 4m class telescopes
by analysing the spectra of two massive stars in the OB\,10 association 
of our nearest giant spiral, 
Andromeda (= M31 = NGC\,224). OB\,10 (van den Bergh 1964, 
Massey et al. 1986) is located at 23.6$'$ from the centre of 
M31, and 
has an apparent size of 2$\times$1.0$'$. Assuming a distance to 
M31 of 783\,kpc (Holland 1998, Stanek \& Garnavich 1998, 
assumed throughout this paper) the association's
projected size is hence $\sim450\times225$\,pc. 
Its stellar content was 
investigated by Massey et al. \shortcite{Mas86} (and later
Massey et al. 1995), who found     
two Wolf-Rayet candidates, OB\,10-WR1 (classified as
WC6-7) and OB\,10-WR2 (classified only as WN) and a late O-type 
supergiant (OB\,10-150; O8.5Ia(f)). The presence of 
such massive stars implies that it is a 
very young association, having formed in the last 
$\sim$5\,Myrs. The de-projected position of OB\,10 would suggest
it has a galactocentric distance of 5.9\,kpc. 
Blair, Kirshner \& Chevalier \shortcite{Bla82} have determined an  
abundance gradient in M31 from a combination 
of observations of H\,{\sc ii} regions and supernova remnants, 
deriving $-0.03 \pm0.01$\,dex\,kpc$^{-1}$. Hence, OB\,10 is expected
to have a metallicity of $\sim$9.1\,dex, a
factor of two greater than the Galactic solar neighbourhood value. 

In this paper, William Hershel Telescope (WHT) spectroscopy of 
OB\,10-64 (B0\,Ia) and OB\,10-WR1 (WC6) are combined to 
reveal the actual
metal abundance of this region of M31, plus the properties of 
WC-type stars in a galaxy similar to our own. 
Designations are taken 
from Massey et al.\  \shortcite{Mas86}; OB\,10--64 has  an alternative 
designation of 41-2265 from Berkhiujsen et al.\ \shortcite{Ber88}, 
while OB\,10--WR1 is also known
as IT5--19 by Moffat \& Shara \shortcite{MS87}. 

\begin{table*}
% \centering
\caption{Observing log and fundamental properties of the M31-OB\,10 targets.
UBV photometry is
taken from Massey et al. (1986), and M$_{\rm V}$ for each star
is calculated assuming 
a distance modulus 24.47$\pm$0.07 (Holland 1998, Stanek \& Garnavich 1998). 
The spectral range quoted for the ISIS data is for the  
unvignetted region; the full blue 
range was not usable due to the large size of the EEV42-80 CCD and the 
small entrance window to the ISIS cameras. The resolution is estimated
from the width of arc lines measured at the adopted slit-width. 
The signal-noise-ratio (S/N) per pixel
is measured in the continuum after binning by 2 pixels in the spectral
direction for the ISIS blue arm, and unbinned data in the red.
Note that only 1.0 hour integration was obtained for the red ISIS arm in Sept
1998. The S/N per pixel in the Keck HIRES data is after binning to 
0.3\AA\,pix$^{-1}$.
}
\label{phot_pars}
\begin{tabular}{l@{\hspace{-2mm}}c
@{\hspace{2mm}}l@{\hspace{2mm}}l@{\hspace{1mm}}c@{\hspace{2mm}}l@{\hspace{2mm}}
l@{\hspace{2mm}}l@{\hspace{2mm}}l@{\hspace{2mm}}l@{\hspace{2mm}}l}\hline
Star          & Epoch & \multicolumn{2}{c}{Spectral
Ranges (Res)}& S/N & Exp. & V & B$-$V & U$-$B & E(B$-$V) & M$_{\rm V}$ \\
              & & \multicolumn{2}{c}{\AA}   & 
& Hours & mag & mag & mag & mag & mag \\
\hline
WHT ISIS data:\\
OB\,10--64      &    Sep 1998 & 3985$-$4725 (0.8) & $\cdots$ & 50 & 4.5 & 18.10$\pm$0.04     
& $-0.08\pm$0.08   & $-$0.97   & 0.18$\pm$0.08 &  $-$6.93$\pm$0.4 \\
OB\,10--WR1     & Sep 1998 & 3985$-$4725 (0.8) & 
5650$-$6400 (1.6) & 5--15 & 5.5 & 19.32$\pm$0.07 & $-0.56\pm$0.08 & 
$-$0.45 & 0.18$\pm$0.08 & $-$5.4$\pm$0.4\\
OB\,10--WR1     & Nov 1998 & 3640$-$6227 (3.5) &
6404$-$9277 (7) & 5-15 & 0.5 & $\cdots$ & $\cdots$ & $\cdots$ & $\cdots$  & 
$\cdots$ \\
Keck~I HIRES data:\\
OB\,10--64 & Oct 1999 & 6530--6630 (0.2) & $\cdots$ & $\sim$60 & 4 & 
$\cdots$ & $\cdots$ & $\cdots$ & $\cdots$  & $\cdots$ \\
HST STIS data:\\
OB\,10--64 & Oct 1999 &  1120--1716 (1.5)   & $\cdots$  & $\sim$20 & 2.3 & 
$\cdots$ & $\cdots$ & $\cdots$ & $\cdots$  & $\cdots$ 
\\\hline	
\end{tabular}

\end{table*}

Our observations are presented in Sect.~\ref{obs}. Since the methods 
used to analyse the spectra of the M31 B-type supergiant and 
Wolf-Rayet star differ considerably, they are discussed separately below in
Sect.~\ref{Bsuper} and Sect.~\ref{WR}, with a general discussion 
presented in Sect.~\ref{discussion}.

\section[]{Observational data and reduction}\label{obs}

WHT spectroscopy was obtained for OB\,10-64 and OB\,10--WR1 during Sep--Nov
1998 as part of a programme to observe early-type stars in the spiral 
galaxies M31 and M33. For OB\,10--64 these data were supplemented 
by Keck~I high resolution echelle spectrograph (HIRES) data around 
the H$\alpha$ region of the spectrum and Hubble Space Telescope 
(HST) Space Telescope Imaging Spectrograph (STIS) UV spectra. 

HD\,167264, a Galactic early B-type supergiant, was further observed in 
September 1998 with exactly the same instrumental setup. For
comparison with OB\,10--WR1, HD\,92809 (WC6) was observed at the Mt Stromlo
\& Siding Spring Observatory (MSSSO) in Dec 1997. Table\,\ref{phot_pars}
summarizes the M31 spectroscopic observations and published photometry.

\subsection{WHT spectroscopy}

The double armed spectrometer ISIS spectroscopy was used to
observe OB\,10-64 and OB\,10--WR1 on the nights of the 
28-29 September 1998. The spectrograph slit was positioned 
so that it included both stars, their separations on 
the sky being only 25$''$. Only the blue arm of ISIS was used
for the first night, with the R1200B grating and a thinned 
EEV42-80 CCD (with format 4096$\times$2048 13.5$\mu$m pixels). 
On the second night, a beam splitting dichroic was used 
to feed both the blue and red channels; for the latter, 
the grating was the R1200R with a TEK CCD used as a detector. 
For the blue spectra, the unvignetted wavelength range was
from approximately 3985\AA\ to 4725\AA\  with a pixel size of 
0.22\AA; the corresponding values for the red spectra were 
5650-6400\AA\ and 0.4\AA. 

Since the above setup was primarily geared towards useful 
photospheric lines for the B-type supergiant, we missed the 
crucial WC abundance lines between 5000--5600\AA. Subsequently,
service observations were taken of OB\,10-WR1 on November 24$^{th}$ 
1998, at a lower but sufficient spectral resolution. The R300B 
and R300R gratings were used in the blue
and red arms at dispersions of 0.9\AA~pix$^{-1}$ and 1.6\AA~pix$^{-1}$ 
covering 3640--6227\AA~and 6404--9277\AA, respectively.

The CCD frames were reduced to wavelength calibrated spectra
using the IRAF reduction system.\footnote{IRAF is written and 
supported by the IRAF programming group at the National Optical 
Astronomy Observatories (NOAO) in Tucson (http://iraf.noao.edu).}
Standard procedures were used to bias correct and flat field 
the stellar images.  There is considerable 
background Balmer line emission from the nebular
region of OB\,10, which can vary
over small spatial scales along the slit. 
Although nebular emission at the sub-pixel level contaminates
our stellar Balmer line observations of 
OB\,10-64, the CCD images showed that the H$\delta$ and H$\gamma$ regions
were less affected than the H$\alpha$~line. 
Further, our spectral resolution is high enough that the important 
wings of H$\delta$ and H$\gamma$ will be unaffected by the background
nebulosity. However, there may be residual errors  
{\em in the cores} of the stellar Balmer line profiles.
The nebular emission around WR1 is particularly intense and 
accurate subtraction proved very difficult, however as shown 
in Figure\,\ref{m31wr1} most of the broader
WR features can be modelled while recognising the narrower nebular
contamination. 
The spectra were wavelength calibrated using CuNe+CuAr lamp 
exposures that interleaved the stellar spectra. Individual
stellar exposures were then cross-correlated to search for
significant wavelength shifts -- none were identified. The
spectra were then combined using {\sc scombine} and a variety
of rejection criteria; the different methods used to combine the data
were found to have little effect on the signal-to-noise and 
the CR rejection success-rate in the final spectra.

\begin{table}
 \centering
  \caption{
Equivalent widths, $W_{\lambda}$, for selected lines of
  He~{\sc ii}, Si~{\sc iii} and Si~{\sc iv} in OB\,10-64 and HD\,167264. 
These lines are very sensitive to $T_{\rm eff}$, and the 
similarity of the line strengths indicate that the effective 
temperatures of the two stars are similar to within 500\,K (see 
Section\,\ref{atmos_param}). Note that the Si\,{\sc iv} line at 
4089 is blended with a nearby O\,{\sc ii} feature, but at this
$T_{\rm eff}$ the Si\,{\sc iv} contribution dominates and the ratio
of the line strengths is approximately 4:1. The equivalent widths
for OB\,10-64 should normally be accurate to 10\%; and for HD167264
accuracies are normally 5\%. }
  \begin{tabular}{llcc}\hline
Species 	& Wavelength (\AA) & \multicolumn{2}{c}{$W_{\lambda}$ (m\AA)}
\\\hline
        	&            	   & OB\,10-64 	& HD\,167264
\\ 
He\,{\sc ii}	& 4541		   & 145	& 110
\\
He\,{\sc ii}	& 4686		   & 290	& 225
\\
Si\,{\sc iii}	& 4552		   & 245	& 225
\\
Si\,{\sc iii}	& 4567		   & 240	& 190
\\
Si\,{\sc iii}	& 4574		   & 120	& 100
\\
Si\,{\sc iv}	& 4089		   & 575	& 590
\\
Si\,{\sc iv}	& 4116		   & 400	& 435
\\
Si\,{\sc iv}	& 4654.14 	   &~\,80 	& ~\,80\\\hline
\label{teff_ew}
\end{tabular}
\end{table}

\begin{figure*}
\epsfig{file=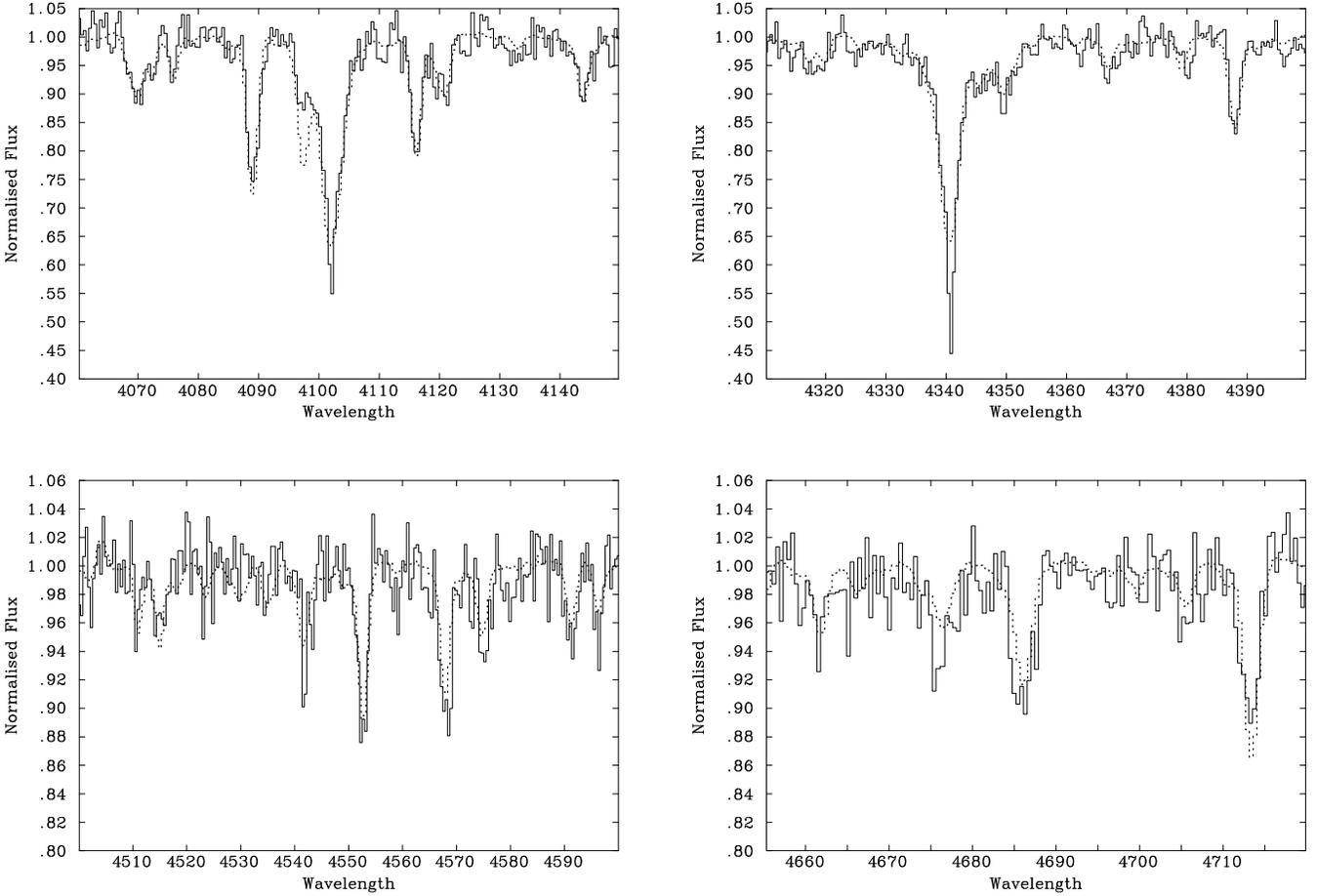,height=18cm,angle=270}
\caption{Selected spectral regions for OB\,10-64 (solid line) and 
HD\,167264 (dotted line), showing temperature sensitive features
(He\,{\sc ii} lines
at 4541 and 4686\AA; Si {\sc iii} lines at 4553, 4568 and 4574\AA\
and Si {\sc iv} lines at 4089 and 4116\AA) and the surface 
gravity diagnostics (Balmer lines: H$\delta$\,4101 and 
H$\gamma$\,4340\AA). The high quality data of this 18$^{m}$ star
allow us to identify and measure relatively weak stellar absorption
lines due to metals and He\,{\sc i}. }
\label{spec_comp}
\end{figure*}

Unfortunately conditions in November 1998 prevented an absolute 
flux calibration for OB\,10--WR1. Nevertheless, a relative calibration 
was achieved, using       the spectrophotometric DO white dwarf standard
star Feige~110. This star was observed for 300 sec with an identical setup,
immediately prior to OB\,10-WR1 and at a similar air mass. % (1.033)
\begin{table}
  \caption{Equivalent widths, $W_{\lambda}$, for metal lines in OB\,10--64
and HD\,167264. Typical errors for OB\,10-64 are 10\% for lines with 
$W_{\lambda}>70$m\AA\, and 20\% for lines with $W_{\lambda}<70$m\AA. 
Measurements of $W_{\lambda}$ for HD\,167264 should normally be 
accurate to 5\%}
  \begin{tabular}{llccl}\hline
Wavelength & $\! \! \!$Species & \multicolumn{2}{c}{$W_{\lambda}$ (m\AA)}
& Status
 \\
~ & ~ & OB\,10-64 & HD\,167264 		
\\\hline
\\
4069.62 & O\,{\sc ii}	& 400 	& 350 	& Blended
\\
4069.89 & O\,{\sc ii}
\\
4070.30 & C\,{\sc iii}
\\
\\
4071.20 & O\,{\sc ii}	& ~\,90 & 100  	& non-LTE
\\
4072.16 & O\,{\sc ii}
\\
\\
4075.85 & C\,{\sc ii}	& 115 	& 145 	& non-LTE
\\
4075.86 & O\,{\sc ii}  
\\
%\\
%4120.28 & O\,{\sc ii}			& Blended
%\\
%4120.58 & O\,{\sc ii}
%\\
%4120.81 & He\,{\sc i}	
%\\
%4120.99 & He\,{\sc i}
%\\
%4121.48 & O\,{\sc ii} 
%\\
%\\
4253.59 & S\,{\sc iii} 	& 115 	& 110	& Blended
\\
4253.74 & O\,{\sc ii}  
\\
4253.98 & O\,{\sc ii} 
\\
\\
4267.02 & C\,{\sc ii} 	& \,80 	& \,75	& non-LTE
\\
4267.27 & C\,{\sc ii} 
\\
\\
4275.52 & O\,{\sc ii}	& 120  	& 110	& non-LTE
\\
4275.90 & O\,{\sc ii}
\\
4276.21 & O\,{\sc ii}
\\
4276.71 & O\,{\sc ii}
\\
4277.40 & O\,{\sc ii}
\\
4277.90 & O\,{\sc ii}
\\
\\
% 4284.51 & N\,{\sc iii} 	& 120  	& 105	& Mis-identified?
%\\
%\\
4317.14 & O\,{\sc ii}  	& \,75 	& \,70 	& non-LTE
\\
\\
4319.63 & O\,{\sc ii}  	& \,85 	& \,80	& non-LTE
\\								   
4319.93 & O\,{\sc ii}  
\\
\\
4349.43 & O\,{\sc ii}  	& 145	& 150	& non-LTE	
\\
\\
4366.89 & O\,{\sc ii}  	& 135	& 100	& non-LTE
\\
\\
4379.05 & N\,{\sc iii} 	& 125	& 110	& LTE
\\
\\
4414.90 & O\,{\sc ii} 	& 110	& 135	& non-LTE
\\
\\
4416.97 & O\,{\sc ii}   & 135	& \,60 	& non-LTE
\\
\\
4481.13 & Mg\,{\sc ii}	& 105 	& \,95  & non-LTE
\\
4481.33 & Mg\,{\sc ii} 
\\
\\
4510.92 & N\,{\sc iii}  & \,55	& \,90	& LTE
\\
\\
4514.89 & N\,{\sc iii}  & \,85	& 120	& LTE
\\
\\
4523.60 & N\,{\sc iii}	& \,40	& \,45 	& LTE
\\
\\
4590.97 & O\,{\sc ii}	& 145 	& \,90	& non-LTE
\\
\\
4595.96 & O\,{\sc ii}	& 100 	& \,70	& non-LTE
\\
4596.17 & O\,{\sc ii} \\\hline
\end{tabular}
\end{table}  

\setcounter{table}{2}
\begin{table}
 \caption[]{continued}
\begin{tabular}{llccl}\hline
Wavelength & $\! \! \!$Species & \multicolumn{2}{c}{Star Number}
& Status
\\
~ & ~ & OB\,10-64 & HD\,167264 
\\\hline
\\
4609.44 & O\,{\sc ii}	& 115 	& \,70 	& non-LTE
\\
4610.14 & O\,{\sc ii}
\\
\\
4630.54 & N\,{\sc ii}	& \,60	& \,55	& non-LTE
\\
4631.38 & Si\,{\sc iv}
\\
\\
4640.64 & N\,{\sc iii}	& 230	& 295	& Blended
\\
4641.82 & O\,{\sc ii}
\\
4641.90 & N\,{\sc iii}
\\
\\
4647.42 & C\,{\sc iii}	& 520 	& 405	& Blended
\\
\\
4649.14 & O\,{\sc ii}	& 875 	& 810	& Blended
\\
4650.25 & C\,{\sc iii}
\\
4650.84 & O\,{\sc ii}
\\
4651.47 & C\,{\sc iii}
\\
\\
4661.63 & O\,{\sc ii}   & \,75 	& \,85 	& non-LTE
\\				  				   
\\   				  
4673.74 & O\,{\sc ii}   & 170	& 135	& non-LTE
\\
4676.24 & O\,{\sc ii} \\\hline
\end{tabular}
\label{OB10_ew}
\end{table}  

\subsection{Keck~I HIRES and HST STIS spectroscopy of OB\,10--64}

OB\,10--64 was observed with the 10m Keck~I telescope, using HIRES
 on 27th September 1997 and 
again on the 6/7 October 1999 giving a total of $4\times3600$s
exposures. A 1.1$''$ slit was used giving a resolution of 
approximately 35\,000, yielding a S/N of $\sim$60 at H$\alpha$
after co-addition and binning in the spectral direction 
to 0.3\AA\,pix$^{-1}$. The CCD echelle spectra were
reduced in the standard manner,
using the same methods as described in McCarthy et al. \shortcite{McC97}.
The most difficult task was removing the strong background H$\alpha$
nebular emission from the stellar spectrum. In fact complete
removal proved impossible, and we have left in the residual 
(negative) nebular H$\alpha$ line in Figs\,\ref{halpha_comparison}
and \ref{ob10_64_hafit}. This led to 
an unrecoverable 0.9\AA~ region of the spectrum, which is however not 
critical to our fitting of the much broader 
stellar H$\alpha$ profile. 

HST spectra were taken with the 
STIS, as part of an extensive Programme (GO7481) 
to observe the UV wind-lines of 
massive blue supergiants in M31 and M33. Two exposures
giving a total of 8500s were taken with the FUV-MAMA detector
on 1st October 1999. The G140L grating and the 52$\times$0.2 arcsec slit
aperture were used. The data were pipelined processed, 
(including wavelength and flux calibration) by the STScI 
On-The-Fly calibration system\footnote{see http://www.stsci.edu/instruments/stis}. 
This processes the data with the 
most up-to-date calibration files, parameters, and software. 
Further details of the observational material can be found in 
Table\,\ref{phot_pars}. 

\begin{table}
\caption{Stellar atmospheric and wind parameters derived for OB\,10--64. }
\begin{tabular}{llll}\hline
Parameter     &  Value \\\hline
$T_{\rm eff}$ & 29000 &      $\pm$ 1000 & (K)  \\
$\log$\,g     & 3.1         & $\pm$ 0.2 &(dex) \\
$\xi$         & 10    &       $\pm$ 5& (\kms)\\
$\dot{M} $    & 1.6$\times10^{-6}$ & $\pm^{0.2}_{0.3}\times10^{-6}$ & (M$_{\odot}$\,yr$^{-1}$) \\ 
$v_{\infty}$  & 1650        & $\pm$70   & (\kms)\\
R$_{\ast}$    & 34          & $\pm$6    &(R$_{\odot})$ \\
$\log$L/L$_{\odot}$ & 5.85  & $\pm$0.24 & (dex) \\
\hline
\label{ob10-64-pars}
\end{tabular}
\end{table}

\subsection{MSSSO spectroscopy}

The Galactic WC6 star
HD\,92809 was observed with the Double Beam Spectrograph (DBS) at 
the 2.3m MSSSO on 1997 Dec 24--27. 
Use of a suitable dichroic permitted simultaneous 
blue and red observations of HD\,92809 covering 3620--6085\AA\ (300B) 
and 6410--8770\AA\ (316R), plus 3240--4480\AA\ (600B) and 8640--11010\AA\ 
(316R). A 2$''$ slit and 
1752$\times$532 pixel SITe CCDs provided a 2 pixel spectral resolution 
of $\sim$5\AA. Wide slit observations were also obtained for HD\,92809
and the HST spectrophotometric standard star HD\,60753 (B3\,IV) to achieve 
reliable flux calibration. Atmospheric correction was achieved by 
observing HR4074 within an air mass of 0.10 from HD\,92809. 
A standard data reduction was again carried out with {\sc iraf}.

These observations were supplemented by high resolution (HIRES),
International Ultraviolet Explorer (IUE), large 
aperture, short-(SWP) and long-wavelength (LWP, LWR) ultraviolet 
observations of HD\,92809. These were combined to provide a single 
high S/N dataset, flux calibrated using the curve obtained by 
Howarth \& Philips \shortcite{HP86}. Finally,
the spectra were transferred to the {\sc starlink} spectrum 
analysis program {\sc dipso} \cite{How98} for subsequent analysis.

\subsection{Photometry, distance and reddening to M31}\label{distance_red}

The published photometry of both stars (from Massey et al. 1986) is
provided in Table~\ref{phot_pars}. Assuming 
(B$-$V)$_{\circ}=-$0.26 for a B0\,Ia atmosphere  \cite{deu76}, 
we estimate an interstellar reddening of
E(B$-$V)=0.18 mag towards OB\,10--64. Massey et al. 
\shortcite{Mas95}
have estimated the mean reddening toward the massive stars
with spectroscopically confirmed types in OB\,10 and
OB8 + OB9, deriving a value of E(B$-$V)=0.16 $\pm$0.02. 
The reddening directly towards OB\,10-64 is 
hence compatible with the mean value for the region. 
Assuming the distance of Holland \shortcite{Hol98} to M31 
of 783$\pm$30\,kpc implies 
M$_{\rm V}=-6.93 \pm0.42$ for OB\,10-64 (from the errors quoted in 
Table\,\ref{phot_pars}). We discuss the implications of 
this in terms of wind and evolutionary models in Section\,\ref{discussion}. 

Our HST STIS data was combined with existing archive FOS and GHRS
data to give a flux distribution from 1120--3300\AA. 
For the value of E(B$-$V)=0.18, the
de-reddened HST/FOS flux distribution for OB\,10--64 matches theoretical
models for $\lambda\ge$2000\AA\, but shows a noticeable drop at shorter
wavelengths, possibly indicating a different far-UV reddening law to the 
Galactic
case  \cite{Sea79}.  Bianchi et al.\,(1996a) investigated the extinction
in M31 using the HST/FOS data for OB\,10-64 and additional M31 OB stars
and concluded that the 2175\AA\ bump is weak or absent in M31. The
slope of the extinction curve was consistent with the Galactic curve, but 
the low reddening of their M31 target stars prevented strong constraints
being placed on the exact shape of this extinction curve.  We have briefly
re-investigated this issue using a slightly different procedure, following
Fitzpatrick (1986) in using other B-type supergiants in the LMC as 
the comparison
stars. We chose Sk--67\,108 (B0\,Ia) and Sk--69\,276 (O9--B0Ia) as the
comparison stars on the basis of the morphology of their UV spectra
compared to OB\,10-64.  These two stars have essentially little or no 
internal LMC extinction and the E(B-V) values 
(0.06 and 0.10 respectively) reflect the reddening from intervening
Galactic material. This Milky Way extinction is similar to what we expect for 
OB\,10-64.  Using these two stars and OB\,10-64 we derive a mean
M31 extinction law, which confirms the lack of a 2175\AA\ feature in M31,
similar to the SMC (Prevot et al.\, 1984). We find a slope for 
M31 which is marginally shallower than the SMC, but steeper than
that of the Milky Way. 
A linear fit of  E($\lambda$-V)/E(B-V) versus ${\lambda}^{-1}$ (${\mu}^{-1}$m) 
gives a slope of approximately 1.8 for
M31; compared with approximately 2.2 for the SMC.
This analysis suggests that the far-UV extinction in M31 is
greater than that predicted by the Galactic extinction law, and that
there is a lack of the 2175\AA~ bump. However this should be treated
with some caution as we have only one star available here, and 
it is only lightly reddened by dust in M31 itself. Ideally 
stars with higher M31 extinction (or preferably a range in 
values) are needed to investigate the relationship fully. 

For the Wolf-Rayet star, broad-band  Johnson photometry are 
(differentially) contaminated by strong emission lines. 
Consequently, the Massey et al. (1986) V-band magnitude for
 OB\,10--WR1 will    overestimate the true continuum flux,
and the observed B--V colour does not permit the determination
of interstellar reddening. Therefore, we assume an identical
reddening to that derived above for OB\,10--64 and
estimate a Smith narrow $v$-band magnitude for OB\,10--WR1 as follows. 
We have convolved our                   WHT spectra of OB\,10--WR1 and 
OB\,10--64 with a Johnson V-band filter,
 revealing $\Delta$V=1.25 mag, in 
excellent agreement with 1.22 from Massey et al.
\shortcite{Mas86}.
Applying the $v$-band filter to the spectrum of OB\,10--WR1 yields 
($v-$V)=0.35 mag, so we estimate $v$=19.67 mag. This agrees
well with the narrow-band 4752\AA\ continuum magnitude of 
19.6 quoted by \cite{MJ98} for OB\,10--WR1. Taking the above
distance and adopting a reddening of $E(B-V)=0.18$
implies M$_{v}=-5.4\pm$0.4 mag. This result is significantly
different from 
the mean absolute magnitudes of $M_{v}=-$3.7 mag and $-$4.8 mag for
respectively, Galactic WC6 and WC7 stars (e.g. Smith et al. 1990).

\subsection{Spectral types}

While spectral typing is often neglected when 
carrying out quantitative atmospheric and wind
analysis of massive stars, we quote our revised types to
allow the stars to be put into context with other 
morphological work. 
Humphreys et al. \shortcite{Hum90} and Massey et al. 
\shortcite{Mas95} gave a spectral type of B1\,I 
for OB\,10--64 based on spectral
data of 2$-$3.5\AA\ resolution and moderate signal-noise, 
while Bianchi et al. \shortcite{Bia96b} 
presented HST ultraviolet spectroscopy
of OB\,10--64, suggesting approximately B0.5\,Ib.
From our higher quality optical and UV 
spectra we have adjusted the spectral type 
slightly to B0\,Ia  (see also Fig.~\ref{ob10_64_uvspec}).
Meanwhile, Massey et al. 
\shortcite{Mas86} classified 
OB\,10--WR1 as a WC6--7 star, which  Moffat
\& Shara \shortcite{MS87} revised 
to WC6+neb!. Our high quality spectrum confirms the WC6 classification
using either the  Smith et al. \shortcite{Smi90} or 
Crowther et al. \shortcite{Cro98} 
schemes, plus the strong nebular contamination.

\section{B-type supergiant analysis}\label{Bsuper}

The non-LTE model atmosphere techniques have been described in detail by
McErlean et al.\ (1999 - hereafter designated as MLD) and here we only 
provide a summary. A grid of non-LTE model atmospheres 
was generated using the code {\sc tlusty} \cite{Hub88}
for effective temperatures, $T_{\rm eff}$, upto 
35\,000\,K and logarithmic gravities from $\log g$\,=\,4.5 down to 
near the Eddington stability limit. Models were calculated
for two helium fractions, viz. $y$~=~0.09 (solar) and $y$~=~0.20, 
where $y$=$N$[He]/$N$[H+He]. The models omit a number of physical 
processes, including  metal line-blanketing  and wind effects. 
Furthermore, the assumption of a plane-parallel geometry 
may be of limited validity for the low gravity objects considered 
here. The consequences of these omissions have been discussed in 
MLD.

The line formation calculations were performed using the codes 
{\small DETAIL} \cite{Gid81} and {\small SURFACE} \cite{But84}. 
Microturbulent velocities, which are 
close to the speed of sound, have previously been found
for B-type supergiants.  Therefore, in the calculation of line
profiles, microturbulence has been included as an extra free parameter.
Metal ion populations and line-profiles were calculated using mainly
the atomic data of Becker \& Butler (1988, 1989, 1990).
Such calculations explicitly include the effects of the relevant ions on
the radiation field. However they do not include the effects of line
and continuum blanketing from other metals. 

Significant difficulties were encountered in running {\small DETAIL} 
particularly for the silicon model ion but also for some other species.
These difficulties occurred mainly at the lowest gravities for effective
temperatures greater than $\sim$ 25\,000\,K. Examination of the line 
profiles indicated that this was caused by either emission or `filling
in' of the profiles. Indeed the {\small DETAIL} calculations normally 
showed an overpopulation of the relevant ionic upper levels coupled with 
large photoionization rates (and subsequent cascades).
MLD postulated 
that the emission was an artefact of their exclusion of line blanketing 
which leads to an overestimate of the UV flux and hence of the 
photoionization rates - further discussion of these problems can 
be found in their paper, while their implications for the current dataset
are discussed below.

\subsection{Atmospheric parameters}\label{atmos_param}

A comparison of the spectra of OB\,10-64 and HD\,167264 indicates that
these stars have very similar, if not identical, atmospheric parameters
within the errors of our methods. This is illustrated
in Fig.~\ref{spec_comp}, where four spectral regions are shown for the two
stars. These regions contain diagnostics of the atmospheric parameters
as discussed below:
\\
\\
{\bf Effective temperature:} As discussed by MLD, both the 
strength of the He\,{\sc ii} features and the relative strengths
of lines due to  Si\,{\sc iii} and Si\,{\sc iv} can be used to
estimate the effective temperature for spectral types near B0. 
The former is an excellent diagnostic with a change of effective 
temperature of 2000K leading to changes in the He\,{\sc ii}
equivalent widths of a factor of two or three; additionally
MLD found that the He\,{\sc ii} line at 4541\AA\ appeared to
be well modelled by the non-LTE calculation. In Fig.~\ref{spec_comp},
two He\,{\sc ii} lines at 4541\AA\ and 4686\AA\ are shown;
their strengths seem very similar in the two stars. This is
confirmed by the observed equivalent widths (see below for
details of the procedures used to estimates these values) 
summarized in Table \ref{teff_ew}.
Although the line strengths are approximately 20-30\% larger for OB\,10-64,
this would correspond to a change in effective temperature of less
than 500K, such is the sensitivity of the line strength to temperature
in this $T_{\rm eff}$ range. 

Lines due to Si\,{\sc iii} (at approximately 4553, 4568 and 4574\AA)
and Si\,{\sc iv} (at 4089 and 4116\AA) are also shown in
Fig.~\ref{spec_comp}. Again the agreement is excellent as is
confirmed by the estimates of the equivalent widths listed in Table 
\ref{teff_ew}. For OB\,10-64, the Si\,{\sc iii} lines are marginally
enhanced and the Si\,{\sc iv} lines marginally weakened compared with
the standard star. This would imply that OB\,10-64 had a lower effective
temperature, in contrast to the He\,{\sc ii} lines that implied that
this star was hotter. However, again the differences are marginal and 
would imply a temperature difference between the two stars of less than
500K.

{\bf Surface gravity:} The profiles of the Balmer lines are useful
diagnostics of the surface gravity and in Fig.~\ref{spec_comp}, the 
H$\gamma$~(at 4340\AA) and H$\delta$~(at 4101\AA) are shown. In the
line wings (which have the greatest sensitivity to the gravity), the
agreement between the two stars is excellent and imply that the
gravities (assuming that the stars have similar effective temperatures)
differ by less than 0.2 dex; the  major uncertainty arises from the
difference in S/N of 
the observational data for OB\,10-64 compared to the bright standard. 
In the line cores, there are significant differences but this 
is likely to be due to 
different amounts of rotational or macroturbulent broadening rather
than a difference in the surface atmospheric parameters. 
Additionally for the OB\,10-64 spectra there were significant amounts 
of background emission from the host galaxy as discussed in the 
previous section; the  discrepancies may in part reflect the  
difficulty in subtracting this background emission accurately.

Hence we conclude that within the observational uncertainties, these
two stars have the same atmospheric parameters. MLD deduced an
effective temperature of 29000K and a logarithmic gravity of
3.1 dex for HD\,167264 from the optical spectroscopy of Lennon et al.\
\shortcite{Len92,Len93}. We have used our better quality ISIS observations 
to check these results using the same diagnostics as MLD, namely
the He\,{\sc ii} lines for effective temperature and the Balmer
line profiles for the gravity. At these relatively high effective
temperatures, the non-LTE calculations of the Si\,{\sc iii} spectrum 
(and too a lesser extent the Si\,{\sc iv} spectrum)
becomes problematic (see MLD for details) and hence these lines
were not be considered.

\begin{figure}
\epsfig{file=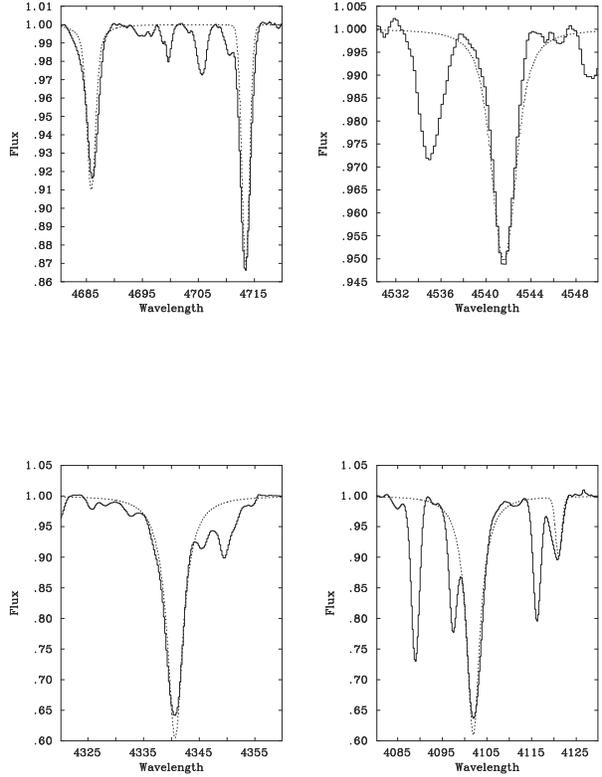,width=8.4cm}
\caption{Observed (solid line) and non-LTE (dotted line) spectra for
HD\,167264. The upper two panels show the He\,{\sc ii} lines at 4541\AA\
and 4686\AA\ which are diagnostics of the effective temperature. 
The lower two panels show the gravity sensitive
Balmer lines, H$\delta$ and H$\gamma$. Excellent agreement is found 
between the line profiles and strengths.}
\label{teff_logg}
\end{figure}

The observed and theoretical spectra are shown in Fig.\ \ref{teff_logg},
with the latter having been convolved with a Gaussian profile with a FWHM of
0.80\AA\ to allow for instrumental, rotational and macroturbulent velocity
broadening. In general the agreement is good, apart from the He\,{\sc ii}
line at 4541\AA, where the wings of the theoretical profile appear to be
too strong. However even for this case, a decrease of only 1000K would lead
to a theoretical profile that is significantly weaker than the observed. 
Hence we conclude that the atmospheric parameters of MLD are reliable 
and adopt them for both HD\,167264 and OB\,10-64. Additionally,
MLD found that a microturbulent velocity of 10 \kms was appropriate for
the Si\,{\sc iii} features in B-type supergiant spectra. They also noted
that the O\,{\sc ii} spectra implied a higher microturbulent velocity --
a result found independently by Vrancken et al.\ \shortcite{Vra99} for
early-B-type giants. Here we adopt a microturbulent velocity of 10 \kms
for both stars but will consider below the effect of a larger value
being appropriate.

\subsection{Chemical composition}

The choice of a standard with atmospheric parameters
effectively identical to those for OB\,10-64 simplifies the
abundance analysis. Our procedure was as follows.
Firstly metal line absorption features were identified in the spectra
of OB\,10-64. For marginal features, the spectrum of HD\,167264
was used as a template and guide. The equivalent widths of the
features were estimated in both stars by non-linear least
squares fitting of single or multiple Gaussian absorption
profiles to the normalised spectra. Normally both the positions
and widths of the Gaussian profiles were allowed to vary. However
for multiplets the relative wavelength separations were fixed,
while for marginal features in OB\,10-64, the widths were set to
the mean value found for well observed lines. The equivalent widths 
are listed in Table \ref{OB10_ew}, apart from those for silicon,
which are in Table \ref{teff_ew}; all estimates have been rounded
to the nearest 5 m\AA.

For most species, these equivalent widths were analysed using the 
non-LTE grids of MLD and the atmospheric parameters discussed above. 
As well as absolute abundance estimates a line-by-line differential
analysis was also undertaken. 
For the N\,{\sc iii} spectrum and one Si\,{\sc iv} line, non-LTE 
calculations were not available
and LTE methods were used; in these cases, the absolute abundance 
estimates should be treated with caution. However, the similarity in 
both the observed line strengths and inferred atmospheric parameters for
OB\,10-64 and HD\,167264 imply that the differential abundance
estimates should be reliable. This was confirmed by re-analysing
the O\,{\sc ii} line spectrum using an LTE technique, where the
{\em differential} abundances agreed with the non-LTE estimates to normally
within 0.01 dex. Both the absolute and differential
abundance estimates are summarized in Table \ref{bsgabund}, together with the
number of features used and whether an LTE or non-LTE formulation was
adopted. Note that the latter information is replicated in Table \ref{OB10_ew},
together with information on which features were included in the
abundance analysis. The error estimates in Table \ref{bsgabund} are unweighted 
standard deviations of the samples; where a significant number of 
features were measured, the uncertainty in the means should be smaller.

\subsection{Photospheric abundances of OB\,10-64}
\begin{figure}
\epsfig{file=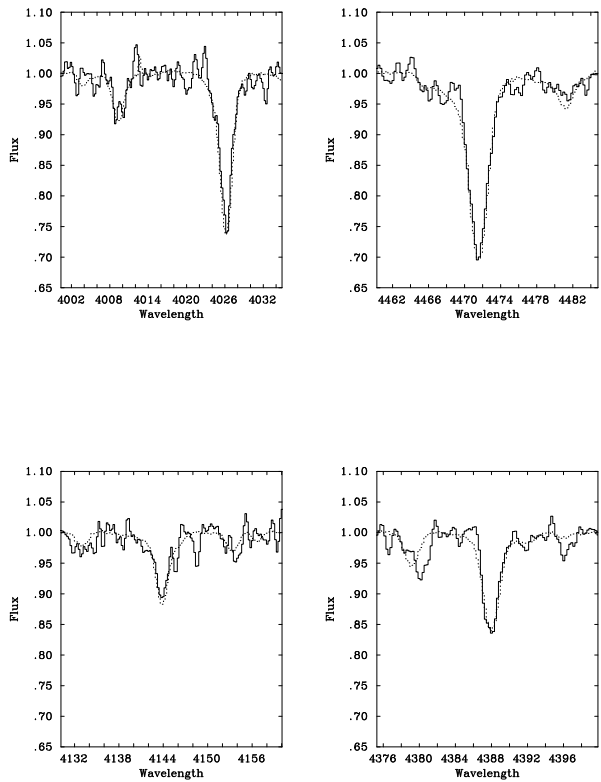,width=8.4cm}
\caption{Selected spectral regions for OB\,10-64 (solid line) and 
HD\,167264 (dotted line), showing He\,{\sc i} diffuse lines
at 4009 and 4026\AA, 4471\AA, 4143\AA, 4387\AA.}
\label{hei_diff}
\end{figure}

The atmospheric parameters and chemical composition of OB\,10-64
appear to be very similar to those of HD\,167264. Below, we discuss
the abundances of individual elements in detail:
\\
\\
{\bf Helium:} The helium abundance in OB\,10-64 has not been determined
explicitly. This was because most of the He I non-diffuse features were
either badly blended with metal lines (e.g. the line at 4121\AA) or weak
(the lines at 4169\AA\ and 4437\AA). The only well observed non-diffuse
feature is the triplet line at 4713\AA~	 and as can be seen from 
Fig.~\ref{spec_comp}, the profiles for the two stars are very similar.
Additionally the moderate quality of the observational data for 
OB\,10-64 would make reliable profile fitting of the helium diffuse 
lines difficult. However in Fig.~\ref{hei_diff}, selected diffuse
helium lines are shown for the two stellar spectra. In general, agreement
is excellent and we conclude that within the observational uncertainties
the helium abundances in the two objects are similar.

{\bf Carbon:} The carbon abundances are based on a single weak feature  
(the 4267\AA\ doublet) which is of similar strength in each star. 
MLD have commented on the fact that the line appears difficult to 
model properly with the non-LTE techniques that we use here. Even in 
B-type dwarfs, calculation of this transition is problematic in both 
LTE and non-LTE.  Hence while the
absolute abundance estimates should be treated with considerable caution, 
the similarity in line strength points to both stars having similar 
C photospheric abundances. 
\\
\\
{\bf Nitrogen:} This element is particularly important as it provides
an excellent diagnostic for the mixing of nucleosynthetically
processed material from the stellar core to the surface. Hence,
a search was made for lines of both N\,{\sc ii} and  N\,{\sc iii}
in the spectrum of OB\,10-64. Unfortunately all the detections
were marginal and the corresponding line strengths should be
treated with caution. The single N\,{\sc ii} line implies that the
two stars have similar nitrogen abundances while the four
N\,{\sc iii} lines implies that OB\,10-64 is depleted by 
approximately 0.2 dex. However it should be noted that the
theoretical N\,{\sc iii} line strengths are very sensitive to the
adopted effective temperature, which coupled with the
observational uncertainties, implies that this difference is probably
not significant.

\begin{table*}
% \centering
  \caption{Absolute and differential abundances for OB\,10-64 and HD\,167264. 
Where more than one feature was measured for a given species (as denoted by
$n = $ number of available lines), the unweighted mean is quoted with the error being the 
standard deviation of the sample. 
The method of line formation calculation is noted, 
and for further information see the text. For reference
we quote the abundances in solar neighbourhood
B-type stars from Gies \& Lambert (1992) for all elements apart from 
Mg which is from Kilian (1994)}
\begin{tabular}{llllrll}\hline
Ion & \multicolumn{2}{c}{[$\frac{X}{H}$]} & $\Delta$[$\frac{X}{H}$] & $n$ & Method
& Solar B-star \\
        	& OB\,10-64 	& HD\,167264 & & & & abundances \\\hline
\\
C\,{\sc ii}	& 7.83		& 7.78		& +0.05		&  1  & non-LTE &  8.20
\\
N\,{\sc ii}	& 7.96		& 7.90		& +0.06		&  1  & non-LTE &  7.81
\\
N\,{\sc iii}	& 8.40$\pm$0.07	& 8.60$\pm$0.27	& -0.20$\pm$0.30 & 4  & LTE     &  7.81
\\
O\,{\sc ii}	& 8.69$\pm$0.34	& 8.62$\pm$0.23	& +0.07$\pm$0.15 & 13 & non-LTE &  8.68
\\
Mg\,{\sc ii}	& 7.36		& 7.30		& +0.06		 &  1 & non-LTE &  7.45
\\
Si\,{\sc iii}	& 7.54$\pm$0.18	& 7.34$\pm$0.12	& +0.20$\pm$0.10 &  3 & LTE     &  7.58
\\
Si\,{\sc iv}	& 7.7:		& 7.8:		& -0.16		 &  2 & see text & 7.58
\\
\hline
\label{bsgabund}
\end{tabular}
\end{table*}
~\\
\\
{\bf Oxygen:} This is the best observed species with 13 lines being
analysed. The oxygen abundances in the two stars are very similar as 
can be seen
from the absolute abundance estimates. It is encouraging that the standard
error for the differential abundance is significantly smaller implying
that the use of a differential method is indeed removing some sources of
systematic error. Assuming that the remaining errors are distributed
normally, the estimated uncertainty in the mean differential
abundance would be reduced to approximately $\pm 0.02$\, and hence
there is marginal evidence for a slight enhancement in OB\,10-64.
\\
\\
{\bf Magnesium:} The estimates are again based on a single feature but 
the two stars appear to have similar abundances.
\\
\\
{\bf Silicon:} The Si\,{\sc iii} lines were analysed using both non-LTE
and LTE methods. The former leads to higher absolute estimates but a
similar differential abundance as for the LTE calculations. As discussed
by MLD, the non-LTE profiles of the Si\,{\sc iii} lines show some emission
in their wings in this temperature regime; this is probably an artifact
of the lack of line blanketing in the calculations. Hence in Table 
\ref{bsgabund}
the LTE results are summarized but the absolute abundance estimates
should be treated with caution. The differential abundance implies 
a small silicon enhancement in OB\,10-64 in contrast to that found 
from the two Si\,{\sc iv} lines. The latter result should also be treated 
with some caution as the analysis of one of the Si\,{\sc iv} lines was 
undertaken
in LTE (it was not included in the non-LTE model ion), which will make
the absolute abundance estimates particularly unreliable. Additionally 
the theoretical line strengths are extremely sensitive to the adopted
effective temperature with, for example, a change of less than 1000K
being sufficient to eliminate the discrepancy. Hence we believe that
the Si\,{\sc iii} differential abundance is the more reliable.

Despite the difficulties for individual ions discussed above, we believe
that the differential abundance estimates should be generally reliable.
This reflects the similar atmospheric parameters and metal line
equivalent widths for the two stars. The former ensures that the method 
of analysis (LTE or non-LTE) is relatively unimportant and indeed we
confirmed this by re-analysing the O\,{\sc ii} equivalent widths
using LTE methods. Additionally the similarity in line strengths
makes the choice of the microturbulent
velocity relatively unimportant provided
it is the same in both stars; for example, re-analysing the metal
line spectra with a microturbulent velocity of 15 km\,s$^{-1}$
changed the differential estimates by less than 0.05 dex.

\begin{figure}
\epsfig{file=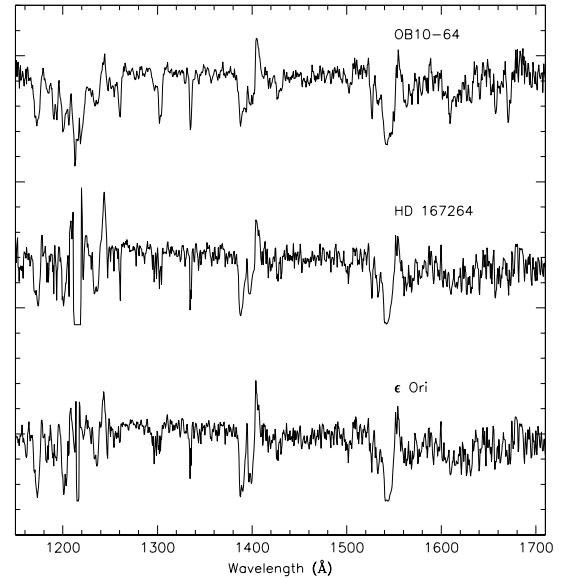 ,width=8.4cm}
\caption{A compilation of galactic IUE spectra of two B0\,Ia
Galactic stars and the STIS spectra of OB\,10-64.}
\label{ob10_64_uvspec}
\end{figure}

\subsection{Wind Analysis and Stellar Wind Momentum}

OB\,10-64 is an ideal object to test the validity of the concept of the
wind momentum -- luminosity relationship. It has a well determined chemical
composition comparable to the solar neighbourhood and an equally
well determined effective temperature and gravity. The reddening is moderate
and the absolute magnitude follows from the distance to M31 (see
Table \ref{phot_pars}). Using a non-LTE
 model atmosphere code with the
corresponding effective temperature and gravity we can calculate the emergent
stellar flux and determine the stellar radius from the absolute V-band
magnitude (see Kudritzki 1980). We obtain a radius of $R = 34 R_{\odot}$.
This means that OB\,10-64 has stellar parameters very similar to the
galactic object HD 37128 ($\epsilon$ Ori), which has been investigated recently
by Kudritzki et al. \shortcite{kud99} 
in their study of the wind momentum -- luminosity
relationship of galactic A- and B-supergiants. We  therefore  expect both
objects to have comparable wind momenta. HD167264, the Galactic comparison
object for the abundance study in this paper, has a very uncertain distance
(and, therefore, radius and luminosity). From the similarity of the spectra
we can only assume that its luminosity must be comparable to OB\,10-64 and
HD 37128.

\begin{figure}
\epsfig{file=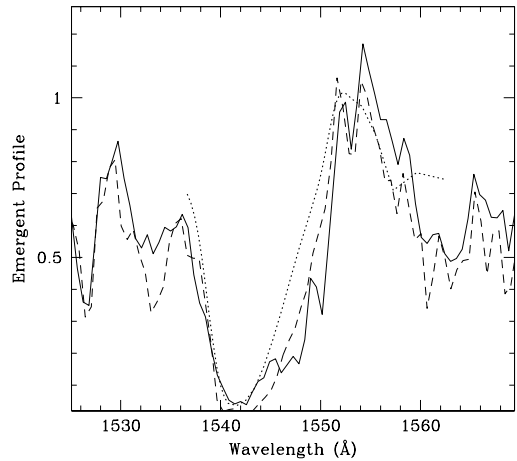,width=8.4cm}
\caption{The wind analysis fit (dotted line)
for $v_{\infty}$ shown for the C\,{\sc iv}
line only. OB\,10--64 is the solid line and HD167264 is the dashed. 
The HD167264 high-resolution IUE spectra has been degraded to the 
STIS resolution. A $v_{\infty}$=1650 \kms is found.
}
\label{uv_finefi}
\end{figure}

\begin{figure}
\epsfig{file=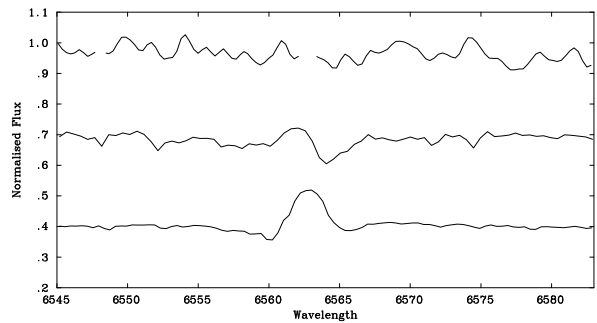 ,width=8.4cm}
\caption{Comparison between the H$\alpha$ profiles 
of OB\,10--64 (upper) and HD167264 (middle) and HD37128 (lower). 
The Keck~I HIRES data of
OB\,10--64 has been convolved with a Gaussian to yield 
the same resolution as the spectra of HD167264 and HD37128 
(see Lennon et al. 1993
for details of Galactic star observations), and re-binned to
a pixel size of 0.3\AA. The nebular lines 
in OB\,10--64 which could not be subtracted 
properly have been snipped out. The profiles of the
three stars are qualitatively very similar indicating
similar wind characteristics. Further, the
differential photospheric analysis is not compromised
by different wind properties of OB\,10--64 and HD167264.} 
\label{halpha_comparison}
\end{figure}

An inspection of Fig.~\ref{ob10_64_uvspec} immediately indicates
(albeit in a qualitative manner)
that the stellar wind properties of the three objects cannot be very 
different.
For the quantitative wind 
analysis of OB\,10-64 we proceed exactly in the same way
as described by  Kudritzki et al. \shortcite{kud99}. 
First, we measure the terminal velocity
$v_{\infty}$ 
of the stellar wind from radiative transfer fits of the strong UV resonance
lines of N\,{\sc v}, C\,{\sc iv} and Si\,{\sc iv}  
using the method developed by Haser \shortcite{has95}
(see also Haser et al. 1995,  Haser et al. 1998, Lamers
et al. 1999). This is 
similar to the technique applied by Howarth et al. 
\shortcite{how97}. The fit yields
$v_{\infty} = 1650$ \kms for both OB\,10-64 and HD167264 
(see Fig.~\ref{uv_finefi}), which is very close to the value of 1600\,\kms
found for HD 37128.

\begin{figure}
\epsfig{file=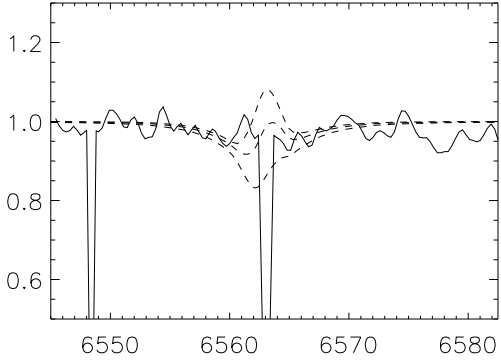 ,width=8.4cm}
\caption{The wind model fit to the Keck~I HIRES spectra
of H$\alpha$ assuming the parameters outlined in
Table\,\ref{ob10-64-pars}. The sharp absorption-like
features reflect the difficulty or reliably subtracting the
background nebular lines (H$\alpha$ and
[N\,{\sc ii}] 6548.1\AA), and these parts of the stellar
spectra are unrecoverable. Despite this, the
H$\alpha$ stellar wind profile can be fitted quite
accurately. Three NLTE unified model calculations for 
$\dot{M}$ = 1.00, 1.50, 1.75 $10^{-6}$
M$_{\odot}$/yr are shown as dashed curves.}
\label{ob10_64_hafit}
\end{figure}

\begin{figure}
\epsfig{file=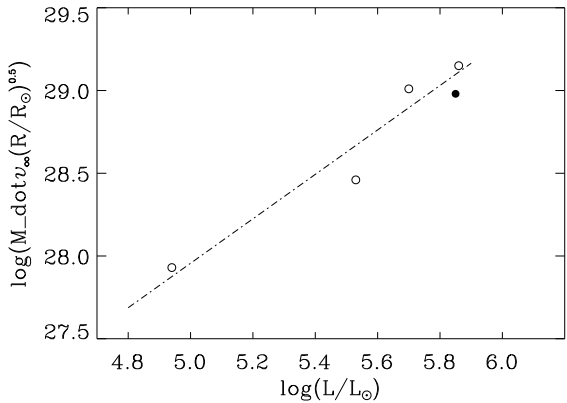 ,width=8.4cm}
\caption{Modified Wind Momentum of OB\,10--64 (solid) compared with the
galactic supergiants of early B-type spectral type in Kudritzki et al. (1999)
(open circles). The
regression curve is a fit to the galactic objects only.} 
\label{ob10_64_mom}
\end{figure}

Subsequently, the mass-loss rate $\dot{M}$ can be obtained from a fit of
the observed line profile of H$ {\alpha}$. This requires a non-LTE  
unified model atmosphere analysis incorporating the
stellar wind and spherical
extension  \cite{kud99}. Assuming a radius of 34 R$_{\odot}$
and a terminal velocity of $v_{\infty} = 1650$ \kms, we calculated a sequence
of unified models with different mass-loss rates. Fig.~\ref{ob10_64_hafit}
gives an idea of the accuracy of the line profile fit, which is
somewhat restricted due to the
central contamination by the H\,{\sc ii}-region emission. 
Despite this, a reasonable determination of the mass-loss rate is certainly
possible. Values as low as 1.0$\times10^{-6}$ M$_{\odot}$yr$^{-1}$ 
can clearly be ruled
out together with values above 1.75$\times10^{-6}$ M$_{\odot}$yr$^{-1}$. 
We adopt 1.6$\times10^{-6}$ M$_{\odot}$yr$^{-1}$
as the model which gives the best compromise 
fit in Fig.~\ref{ob10_64_hafit}. This means that OB\,10-64
has a slightly lower  mass-loss rate than HD 37128
(2.5$\times10^{-6}$ M$_{\odot}$yr$^{-1}$, Kudritzki et al. 1999). 
The modified stellar
wind momentum log$(\dot{M}v_{\infty}(R/R_{\odot})^{0.5})$,
which is
expected to be a function of luminosity only, is 28.98 (in cgs-units)
compared to 29.15 for HD 37128. Fig.~\ref{ob10_64_mom} compares the modified
wind momenta of OB\,10-64 and of all the galactic early B-supergiants
studied by Kudritzki et al. \shortcite{kud99}.   
Within the intrinsic scatter of the wind
momentum -- luminosity relationship the agreement is excellent. This
is the first direct demonstration that the concept of the WLR works
for objects of this spectral type (i.e. early B-types)
outside our own Galaxy.

\section{Wolf-Rayet analysis}\label{WR}

\subsection{Technique}

We use the non-LTE 
code of Hillier \& Miller \shortcite{HM98} 
which iteratively solves the transfer equation in the co-moving frame 
subject to statistical and radiative equilibria in an expanding, spherically 
symmetric and steady-state atmosphere. Relative to earlier versions of this code (Hillier 1987, 1990), two major enhancements 
have been incorporated, of particular relevance to WC-type stars, namely (i)
line blanketing, (ii) clumping.  

Our analysis of OB\,10-WR1 follows a
similar method to that recently applied to Galactic WC stars 
by Hillier \& Miller \shortcite{HM99}. Specific details of the (extremely
complex) model atoms used here for our quantitative analysis are 
provided in Dessart et al.\ \shortcite{Des00}, together with
information on the source of atomic data used for He\,{\sc i-ii}, 
C\,{\sc ii-iv}, O\,{\sc ii-vi}, Si\,{\sc iv} and Fe\,{\sc iv-vi}.
We use an identical (slow) form of the velocity law, such that 
acceleration is modest at small radii, but continues to large distances.
We assume that the wind is clumped with a volume filling factor, $f$, and that
there is no inter clump material. Since radiation instabilities are not 
expected
to be important in the inner wind we parameterise the filling factor 
so that it approaches unity at small velocities. 
Clumped and non-clumped spectra are very similar, except that line profiles 
are slightly narrower with weaker electron 
scattering wings in the former. Although 
non-clumped models can be easily rejected \cite{HM99}, 
because of the severe line blending in WC winds 
$\dot{M}/\sqrt{f}$ is derived by our spectroscopic analysis, rather 
than $\dot{M}$ and $f$ independently.

\begin{figure*}
\epsfxsize=18cm \epsfbox[20 220 530 470]{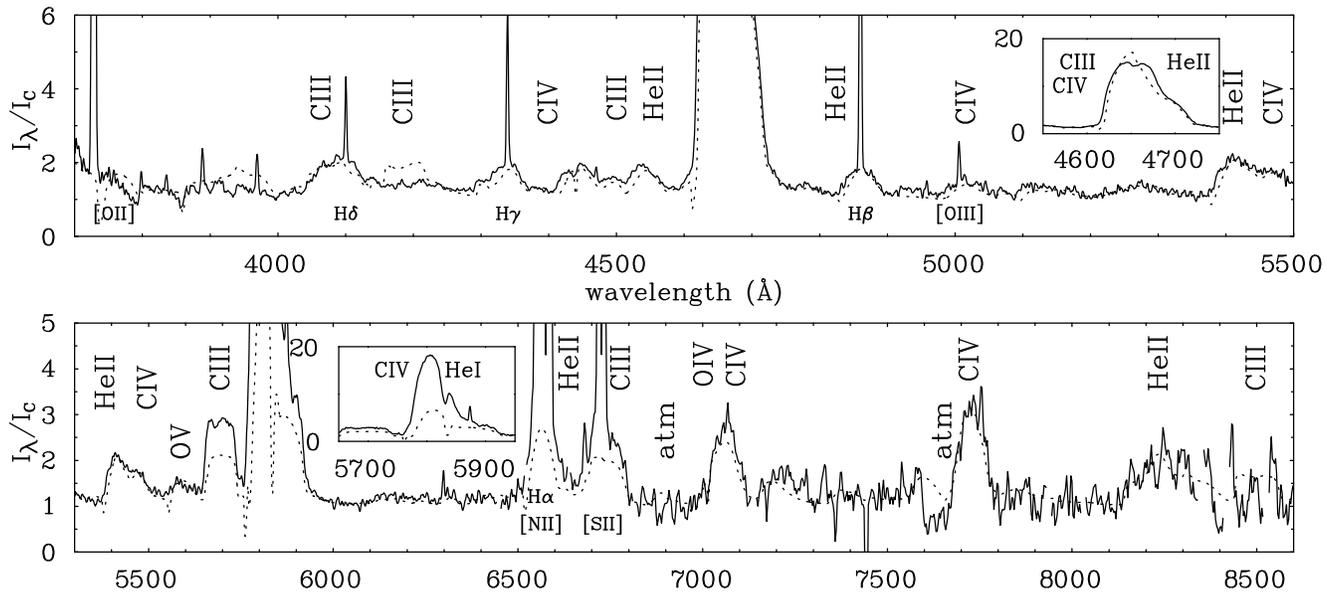}
\caption{
Comparison between rectified WHT/ISIS spectroscopy of 
M31-OB\,10--WR1 (solid) and our synthetic model (dotted). Strong nebular 
emission lines severely contaminate the He\,{\sc ii} 
$\lambda$6560 and C\,{\sc iii} $\lambda$6740 emission lines,
while atmospheric features in the far-red have not been removed.
% Derived parameters are $T_{\ast}$=75kK, log$(L/L_{\odot})$=5.7, 
% $\dot{M}/\sqrt{f} \sim 10^{-4.3} M_{\odot}$yr$^{-1}$, 
% $v_{\infty} \sim3000$ km\,s$^{-1}$, and C/He$\sim$0.1 by number (C/O=4
% is assumed).
}
\label{m31wr1}
\end{figure*}

\begin{figure*}
\epsfxsize=18cm \epsfbox[20 340 530 700]{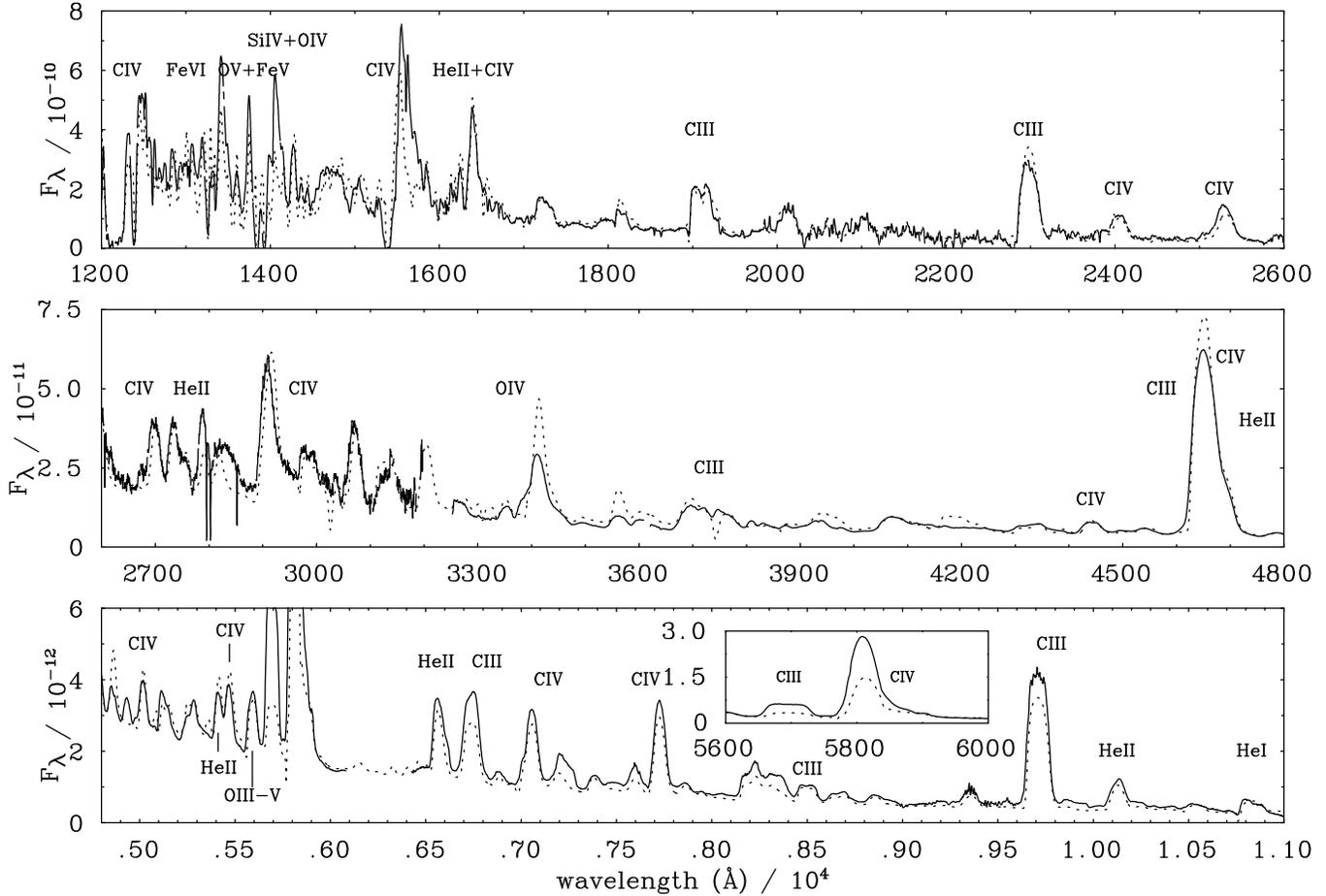}
\caption{
Comparison between de-reddened (E$(B-V)$=0.50 mag)
IUE/HIRES and MSSSO/DBS
spectroscopy of HD\,92809 (solid) with our synthetic model (dotted).
% Derived parameters are $T_{\ast}$=75kK, log$L/L_{\odot}$=5.3, 
% $\dot{M}/\sqrt{f} \sim10^{-4.8} M_{\odot} {\rm yr}^{-1}$, 
% $v_{\infty}$=2280 km\,s$^{-1}$, and C/He$\sim$0.3 by number (C/O=4
% is assumed).
}
\label{wr23}
\end{figure*}

\subsection{Spectroscopic analysis of OB\,10--WR1}

As usual, a series of models were calculated in which stellar 
parameters ($T_{\ast}$, log\,$L/L_{\odot}$, $v_{\infty}$
$\dot{M}/\sqrt{f}$, C/He, O/He) were adjusted until 
the observed ionization balance, line strengths, widths, 
and absolute $v-$band flux were reproduced. 
Because of the (substantial) effect 
that differing mass-loss rates,  temperatures and elemental abundances 
have  on the emergent spectrum, this was an iterative process.

The wind ionization balance is ideally selected on the basis of isolated
optical lines from adjacent ionization stages of carbon (e.g. C\,{\sc iii}
$\lambda$6740/C\,{\sc iv} $\lambda$7700) and/or helium (He\,{\sc i} 
$\lambda$5876/He\,{\sc ii} $\lambda$5412). In practice, this was 
difficult to achieve because of the severe blending in WC winds, so our
derived temperature should be treated as approximate.  Current 
spectroscopic models fail to reproduce the absolute and relative
strength of the WC classification lines at C\,{\sc iii} $\lambda$5696 and 
C\,{\sc iv} $\lambda\lambda$5801-12 (see Hillier \& Miller 1998 
and Dessart et al. 2000).
As in other recent spectroscopic studies of WC stars, He\,{\sc ii} 
$\lambda$5412/C\,{\sc iv} $\lambda$5471 were selected as the principal
C/He diagnostic since the relative strength of  these features are 
insensitive to differences of temperature or mass-loss rate. 
Oxygen abundances were impossible to constrain, since the 
principal diagnostic region spans $\lambda\lambda$2900--3500 
\cite{HM99}. Consequently, we adopt C/O$\sim$4 by number as
predicted by stellar evolutionary models (Schaller et al. 1992).
   Based on results
obtained for OB\,10--64 in Sect.~\ref{Bsuper}, we shall adopt 
solar abundances for Si and Fe.

As discussed by Hillier \& Miller \shortcite{HM98}, 
many weak spectral features sit upon
the continuum so that ideally one would wish to compare synthetic
models with fluxed spectroscopy. Since absolute fluxes are unavailable
for OB\,10--WR1, we have instead rectified the dataset, taking care
to use as few `continuum' regions as possible. 

Our ISIS spectroscopic data
 of OB\,10--WR1 is  shown in Fig.~\ref{m31wr1}.
Overall, the spectrum is reasonably well reproduced by our model,
except that C\,{\sc iii} $\lambda$5696 and C\,{\sc iv} 
$\lambda\lambda$5801-12 are too weak. Nebular emission lines
are prominent in the observed spectrum, and contaminate 
the He\,{\sc ii} $\lambda$6560 and C\,{\sc iii} $\lambda$6740 stellar
lines. We find $T_{\ast}\sim$75 kK, log$L/L_{\odot}$=5.68$\pm$0.16, 
$v_{\infty}\sim 3000$ km\,s$^{-1}$ and 
$\dot{M}/\sqrt{f} \sim10^{-4.3} M_{\odot} {\rm yr}^{-1}$, with
C/He$\sim$0.1 by number, from He\,{\sc ii} 
$\lambda$5412/C\,{\sc iv} $\lambda$5471, despite the low S/N achieved
in this region from a single 30 minute ISIS       exposure. 
The wind 
performance number of OB\,10--WR1 is $\dot{M} v_{\infty} / (L/c)
\sim$14 for our assumed $f$=0.1. 

In order to better constrain the stellar parameters of OB\,10--WR1,
one would require UV spectroscopy with HST/STIS to permit a 
reliable reddening determination (the chief source of uncertainty in 
log$L/L_{\odot}$), $v_{\infty}$ via P Cyg resonance lines, and 
oxygen abundances from the near-IR diagnostics.
For E$(B-V)$=0.18, the expected UV continuum flux from OB\,10--WR1 
is 2$\times 10^{-16}$ erg\,cm$^{-2}$\,s$^{-1}$\,\AA$^{-1}$ at 3000\AA, 
or 1$\times 10^{-15}$ erg\,cm$^{-2}$\,s$^{-1}$\,\AA$^{-1}$ at 1500\AA,
assuming a standard Galactic extinction towards OB\,10 (though see 
Sect.~\ref{distance_red}).

\subsection{Quantitative comparison with HD\,92809 (WC6)}\label{hd92809}

How do the   parameters of OB\,10--WR1 compare with WCE
counterparts in other galaxies? 
Amongst Galactic WC6 stars, only HD\,92809 (WR23) and
HD\,76536 (WR14) are moderately reddened and lie at 
established distances \cite{KH95}.  Since high 
quality flux calibrated spectroscopy of HD\,92809 are
available to us (Sect.~\ref{obs}), we now carry out 
an identical analysis of it, permitting a   
quantitative comparison to be carried out.
 Although numerous Galactic WCE
stars have been studied by Koesterke \& Hamann \shortcite{KH95}, 
including HD\,92809, the inclusion of 
line blanketing and clumping into models
has a major effect on the 
stellar luminosity and wind density, justifying the need for a
re-evaluation of its properties 
\cite{Des00}. 

\begin{table}
  \caption{Emission equivalent widths ($W_{\lambda}$ in \AA) for 
selected lines in the optical spectra of OB\,10--WR1 and HD\,92809, with
typical errors of $\pm$10\%. More 
uncertain values for OB\,10--WR1 are shown in parenthesis when strong
nebular emission severely contaminates the stellar spectrum}
  \begin{tabular}{llcc}\hline
\multicolumn{2}{c}{Spectral feature} & 
\multicolumn{2}{c}{$W_{\lambda}$ (\AA)} \\
Major & Minor & OB\,10-WR1 & HD\,92809 \\ 
\hline
C\,{\sc iii} $\lambda\lambda$4647--50 & He\,{\sc ii} $\lambda$4686 & 813 & 881   \\
He\,{\sc ii} $\lambda$5412 &                           &  53 & 30    \\
C\,{\sc iv}  $\lambda$5471 &                           &   32 & 40    \\
O\,{\sc v}   $\lambda\lambda$5572-5604 & O\,{\sc iii} $\lambda$5590 &   35 & 45   \\
C\,{\sc iii} $\lambda$5696 &                           &  145 &  198  \\
C\,{\sc iv} $\lambda\lambda$5801-12  & He\,{\sc i} $\lambda$5876 & 1085 &  907  \\
He\,{\sc ii} $\lambda$6560 & C\,{\sc iv} $\lambda$6560 & (130)  & 150  \\
C\,{\sc iii} $\lambda\lambda$6727-60 &                           & (100) & 215  \\
C\,{\sc iv} $\lambda$7063  & O\,{\sc iv} $\lambda\lambda$7032-53 &  135 & 140 \\
C\,{\sc iv} $\lambda$7724  & C\,{\sc iv} $\lambda$7737 &  165 & 190 \\
\hline
\label{WR_ew}
\end{tabular}
\end{table}

HD\,92809 lies in the Car~OB1 association, at a distance of
2.63\,kpc, with a narrow $v$-band Smith magnitude of 9.71 mag. 
Our spectral synthesis of HD\,92809, using methods identical
to those described above, indicates a reddening 
of E(B$-$V)=0.50 mag (Koesterke \& Hamann \shortcite{KH95} 
obtained E(B$-$V)=0.38),
such that M$_{v}=-4.08$ mag, substantially lower
than OB\,10--WR1 (M$_{v}=-5.4\pm0.4$ mag).
We adopt $v_{\infty}$=2280 km\,s$^{-1}$ for HD\,92809
following the UV analysis by Prinja et al. \shortcite{PBH90}, 
also somewhat lower than OB\,10--WR1.
The measured emission equivalent widths of 
selected lines in the optical spectra of OB\,10--WR1
and HD\,92809 are listed in Table~\ref{WR_ew}. In general,
HD\,92809 has similar line equivalent widths to 
OB\,10--WR1, 
except that the He\,{\sc ii} $\lambda$5412/C\,{\sc iv} $\lambda$5471 ratio
is much smaller, suggesting a higher carbon abundance.
Single WC stars appear to show remarkably uniform emission equivalent
widths at each spectral type, oblivious of their stellar parameters or
chemistry. For example, Conti \& Massey \cite{CM89}
measured
C\,{\sc iv} $\lambda$5801-12 equivalent widths in the range  440$-$1150\AA\ for
(apparently) single Galactic WC6 stars. Therefore, we can have 
confidence that OB\,10$-$WR1 is also single.

Our spectroscopic analysis was repeated for HD\,92809, with our final
UV and optical synthetic spectrum    compared with observations 
of HD\,92809 in Fig.~\ref{wr23} (de-reddened by
E(B$-$V)=0.50 mag following the reddening law of Seaton \shortcite{Sea79}).
Overall, the comparison is
excellent for line strengths, shapes, as is the match to the 
continuum distribution throughout the spectrum. The number of 
line features that are poorly reproduced is small, 
and again includes C\,{\sc iii} $\lambda$5696, C\,{\sc iv}
$\lambda\lambda$5801--12. 

For HD\,92089, we derive the following stellar parameters:
$T_{\ast}$=75 kK, log$L/L_{\odot}$=5.3, 
and $\dot{M}/\sqrt{f} \sim10^{-4.8} M_{\odot} {\rm yr}^{-1}$.
Fits to He\,{\sc ii} $\lambda$5412 and C\,{\sc iv} $\lambda$5471 
imply C/He$\sim$0.3 by number, confirming the higher carbon abundance
expected from their line ratio. Near-UV oxygen diagnostics are well
matched for HD\,92809 with an assumed C/O=4 number ratio.
As discussed previously by Hillier \& Miller \shortcite{HM99} and 
Dessart et al. \shortcite{Des00}, the inclusion of line blanketing     
into WC models significantly increases stellar luminosities 
relative to previous studies -- taking the higher reddening into 
consideration,
blanketing increases the luminosity for HD\,92809 by $\sim$0.25 dex 
relative to  Koesterke \& Hamann  \shortcite{KH95}. We confirm the
previously derived stellar temperature and carbon abundance for HD\,92809.
A clumping factor of $f$=0.1
reduces the global mass-loss rate by $\sim$3, so that the overall effect of
our study is to decrease the
wind performance number for HD\,92809 to 8 
from 108 (Koesterke \& Hamann 1995).  This dramatically alleviates 
the Wolf-Rayet 
`momentum problem', such that multiple scattering of photons
in their highly stratified atmospheres {\it may} be able to account 
for their mass-loss properties via radiatively driven wind theory.

\begin{table}
  \caption{Comparison between stellar parameters of OB\,10--WR1  in M31
with the Galactic WC6 star
HD\,92809 (WR23). Assuming M$_v$ is appropriate, 
typical uncertainties are $\pm$10kK, $\pm$0.12 dex, 
$\pm$100\,\kms, $\pm$0.05 dex, $\pm$15\% respectively for each parameter. 
 }
\begin{tabular}{l@{\hspace{2mm}}l@{\hspace{2mm}}l@{\hspace{2mm}}c
@{\hspace{2mm}}c@{\hspace{2mm}}c@{\hspace{2mm}}c}\hline
Star & $T_{\ast}$ &$\log L$& $v_{\infty}$&log
$\dot{M}$&C/He&M$_{v}$ \\
     & kK         & $L_{\odot}$&km\,s$^{-1}$&$M_{\odot}$yr$^{-1}$& & mag\\
\hline
OB\,10--WR1 & 75 & 5.7 & 3000 & $-$4.3 & 0.10 & $-$5.4 \\ % m31wr1k R=4.1
HD\,92809   & 75 & 5.3  & 2280 & $-$4.8 & 0.30 & $-$4.1 \\
%WR146      & 57 & 5.7  & 2700 & $-$4.5 & 0.08 & $-$5.3 \\
\hline
\label{WCcompare}
\end{tabular}
\end{table}

\section{Discussion}\label{discussion}

\subsection{Comparison of the Galactic and M31 B-supergiants}
\label{discuss_bsg}
The differential analysis of the B-types
implies that the two stars appear to 
have very similar chemical compositions. There is some
marginal evidence for a small metal enhancement in OB\,10-64 but this
could, for example, just be an artifact of an overestimation of
its equivalent widths from the lower quality spectroscopic data.
We believe the differential oxygen abundance to be particularly
sound given the 13 features used in the analysis and the relatively
small spread in their results. As these are luminous 
supergiants one must qualify the use of absolute abundance values
in any external comparison.  
For oxygen, magnesium and silicon however the abundances
derived in HD167264 
(distance from Sun $\sim$1.7\,kpc; Hill, Walker \& Yang 1986)
are similar to the absolute values
derived in solar-neighbourhood B-type main-sequence stars.
The analysis and absolute abundance determination of the 
dwarfs is more reliable and has been shown to be in excellent 
agreement with H\,{\sc ii} regions and the ISM within $\sim$500\,pc
of the Sun (Gies \& Lambert 1992, Rolleston et al. 2000)
Hence we can be reasonably confident 
that this agreement means the O, Mg and Si 
absolute abundances of the HD167264 and OB\,10-64 can be used.
It appears that the OB\,10 cluster was born from an 
ISM with chemical composition very similar to that of the 
solar neighbourhood, and given its proximity to the centre
of M31 this is a surprising result
(see Fig.\,\ref{oxygen_gradient} and Section\,\ref{chemical_comp}). 

The similar nitrogen and carbon abundances in the two stars are 
interesting. MLD in their study of galactic supergiants 
identified three stellar groupings
that they designated as `normal', `processed' and `highly processed'
on the basis of their CNO line strengths. By `processed' they 
meant stars which have C and N line strength ratios which implied
that their photospheres had been contaminated by the products
of core CN-cyle burning. 
HD\,167264 was classified
as `processed' as it had strong nitrogen lines in comparison
with those of carbon. Assuming that the
designations of MLD are valid, we conclude that OB\,10-64 should
also be classified as `processed' and hence at least its carbon
and nitrogen abundances may not be representative of its natal
composition.

In Section\,\ref{distance_red} we have assumed a distance to 
M31 of 783\,kpc and hence have an estimate of 
M$_{\rm V} = -6.93 \pm0.4$. The model atmosphere analysis
produces a flux distribution for the stellar parameters 
shown in Table\,\ref{ob10-64-pars}, which provides a 
bolometric correction of $-$2.9. Hence we determine an
M$_{\rm bol} = -9.83 \pm0.6$, and 
$\log L/L_{\odot} = 5.85 \pm0.24$. This is typical 
of B-type supergiants (and their possible A-type descendants) 
as found by Kudritzki et al. \shortcite{kud99}, and as 
illustrated
in Fig\,\ref{ob10_64_mom}. 
A comparison
with evolutionary models (see Fig\,\ref{evol_diag}; Schaller
et al. 1992) implies an initial main-sequence
mass for the star of approximately 55M$_{\odot}$. 
Whether the star has just come directly from the 
main-sequence or is a blue-loop star with a 
previous history as a red supergiant is unclear. The Schaller et al. 
models predict that the surface CNO abundances should be drastically
altered if the star has been a red supergiant and is in a blue-loop. 
While we suggest that it may show some evidence of CNO processing, the
quantitative values we derive do not appear to support the star 
having {\em highly} processed core material at its surface indicative 
of a red supergiant origin (where the nitrogen
abundance is predicted to be approximately +1.0\,dex higher than
normal). Whatever
its history, its lifetime is
likely to have been of the order 3.5$\pm$1\,Myrs 
(from  Schaller et al. 1992),
in excellent agreement with the age of the more
evolved WR1 star (with age $3.6\pm0.9$\,Myr, 
see Section\,\ref{discussion_wr}) suggesting 
coeval formation in the OB\,10 association.

\begin{figure}
\epsfig{file=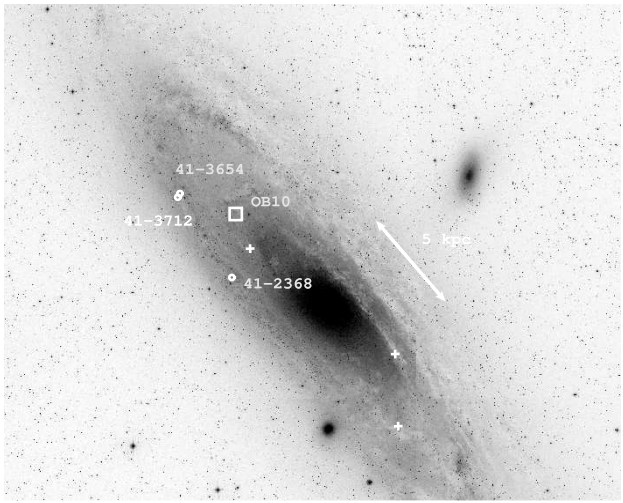,width=8.4cm}
\caption{A mosaic of DSS images of M31 tiled together (hence
the apparent break in the background), with the position 
of the OB\,10 association labelled as the square. For comparison, 
the positions of
three A-F type
supergiants analysed by Venn et al. (2000) are shown, with their other star
A-207 lying off this field-of-view in the outer south west corner of M31. 
North is up, east to the left and the size of the image is 115'$\times$90'.
The three crosses indicate the positions of the three inner disk
H\,{\sc ii} regions analysed by Blair et al. (1982; regions BA75, BA423
and BA289).
The length of the double arrow shows 5\,kpc
at the distance of M31, along the major axis. 
The projected galactocentric radius of the OB\,10 association is 5.9\,kpc. }
\label{m31image}
\end{figure}

\begin{figure}
\epsfig{file=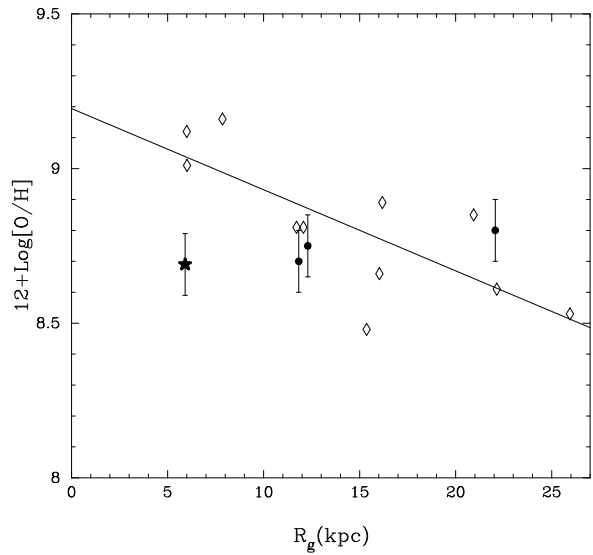,width=8.4cm}
\caption{Oxygen abundances 
as a function of distance from the galaxy centre 
derived in a study of H\,{\sc ii} regions and the 
stars so far analysed in M31.
The open diamonds 
are the nebular H\,{\sc ii} region abundances from 
Blair et al. (1982), and the line shows the 
least-squares-fit to these results, giving 
a gradient of $-0.03 \pm 0.01$~dex\,kpc$^{-1}$. 
 We have recalculated the projected galactocentric
 distances to the H\,{\sc ii} regions
 using an updated distance to M31 of 
 783\,kpc. The solid circles are the 
oxygen results from Venn et al. (2000), from 
the three A-F type supergiants shown in 
Fig.~6. The solid 
star is our photospheric result from OB\,10--64, 
with the error bar representing the standard
error in the mean.}
% (i.e. $\sigma/\sqrt n$, where $n=13$ for oxygen).
\label{oxygen_gradient}
\end{figure}

\subsection{Comparison between Galactic, LMC and M31 WC stars}
\label{discussion_wr}

Table~\ref{WCcompare} reveals that the stellar and chemical 
properties of OB\,10--WR1 and HD\,92809 differ substantially,
despite originating from very similar environments. Following 
the mass-luminosity relation 
for  hydrogen-free WR stars of Schaerer \& Maeder \shortcite{SM92}, 
a present mass of 
18$\pm4 M_{\odot}$ is implied. In contrast, the stellar luminosity of 
HD\,92809 is typical of Galactic WCE stars with a present 
mass of $\sim$11$M_{\odot}$.  Depending on the choice 
of evolutionary model, an age within the range 2.7--4.5Myr is 
implied for  OB\,10--WR1.

Since OB\,10--WR1 is visually the brightest (apparently single)
 WCE star in
M31 (Massey \& Johnson 1998), it is valid to question whether photometry/
spectroscopy of this star
is contaminated by bright stars along similar sight lines, particularly
since 1$''$ corresponds to a physical scale of $\sim$4\,pc at the distance 
of M31.  Excess continuum light would serve to dilute the equivalent 
widths of the WC emission line spectrum, yet            
                          from Table~\ref{WR_ew}, the emission line 
strengths of OB\,10-WR1 agree well with the              single WC6
star HD\,92809. Therefore, in the absence of high spatial resolution imaging
we can have confidence that the observed continuum flux 
is principally that of the WC star, such that its stellar luminosity has
not been artificially enhanced.

 Galactic WCE stars are known to span a range of carbon abundances
and luminosity -- according to Koesterke \& Hamann (1995) 0.1$\le$ C/He 
$\le$0.5, and 5.0 $\le \log L/L_{\odot} \le$ 5.4 for stars with established
distances, and applying a global scaling of 0.3 dex to account for blanketing
(see Sect.~\ref{hd92809}). Therefore, OB\,10--WR1 is amongst the least enriched 
in carbon, with the highest luminosity. These characteristics are reminiscent 
of the nearby WC5+OB binary WR146, recently analysed by Dessart et al. 
(2000) who derived log $L/L_{\odot}$= 5.7 and C/He=0.08 by number. 
Extending the comparison to the LMC WCE stars (all WC4 spectral type), 
reveals quite different conclusions. In contrast with the Galactic sample,
the luminosity of OB\,10--WR1 is fairly typical of LMC WC4 stars, spanning
5.4 $\le \log L/L_{\odot} \le$ 6.0, with comparable C/He abundances established
for Brey\,10 and Brey\,50 (see Fig.4 in Crowther et al. 2000). Therefore,
OB\,10--WR1 is more typical of WCE stars at relatively low metallicity.
A twice solar metallicity for OB\,10 --  inferred from H\,{\sc ii} studies 
(Blair et al. 1982) -- would be very difficult  to reconcile with our 
derived properties of OB\,10--WR1.
                                                                              
\begin{figure}
\epsfig{file=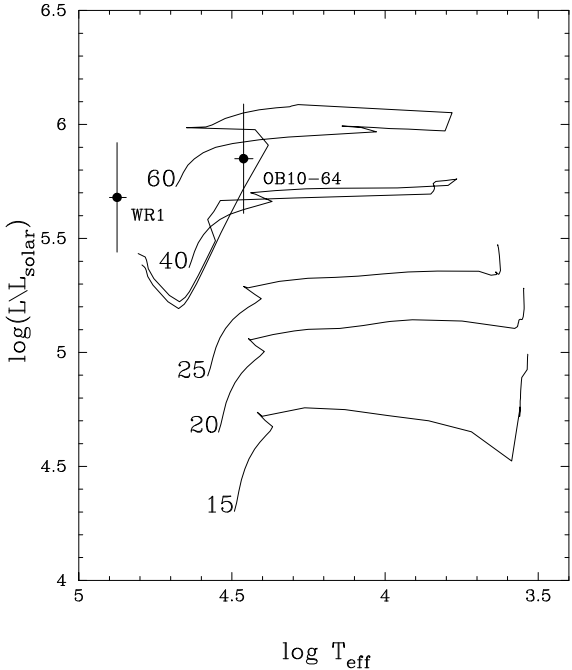,width=8.4cm}
\caption{The evolutionary tracks from Schaller et al. (1992)
are plotted for 15$-$60M$_{\odot}$ stars, as labeled. 
The positions of OB\,10-64, and WR1 are shown. OB\,10 appears
to have evolved from $\sim$36M$_{\odot}$ star, whereas 
WR1 probably evolved from a more massive object. Given the
existence of an evolved star of lower mass than WR1, 
we can infer limits on the time duration of massive 
star formation in this cluster. }
\label{evol_diag}
\end{figure}

\subsection{Probing the chemical composition and evolution
of M31}
\label{chemical_comp}

%As discussed in Section\,\ref{discuss_bsg} the 
%differential oxygen abundance in OB\,10-64 should be reliable
%in the sense that it is similar to the value in local Galactic
%ISM. This star is clearly not particularly metal rich in 
%oxygen, silicon or magnesium. 
In Figure\,\ref{oxygen_gradient} 
we have plotted the oxygen H\,{\sc ii} region abundances from 
Blair et al. (1982) showing the abundance gradient they 
derive, and the position of OB\,10-64. We have recalculated the 
de-projected galactocentric distances of all the objects in 
this plot by assuming a distance of 783\,kpc, an inclination 
angle of 77.5$^{\circ}$, major axis angle of 37.5$^{\circ}$, 
and centre of M31 to be 
$\rmn{RA}(2000)=00^{\rmn{h}}~42^{\rmn{m}}~44\fs4$,
$\rmn{Dec.}~(2000)=41^\circ~16'~08\farcs 97$ (the latter from
Hjellming \& Smarr 1982). The galactocentric distance of OB\,10 is 
5.9\,kpc, and the Blair et al. results suggest that 
the gas metallicity at this position should be significantly above
solar ($\sim$9.1\,dex). Our absolute abundance of 8.7\,dex is 
well below this value, and while we acknowledge the difficulty
in comparing absolute abundances from different methods we are 
confident that OB\,10-64 does not have a metallicity significantly
higher than this value $-$ the star has metal-line strengths and atmospheric
parameters very similar to those of the Galactic standard HD\,167264
as discussed in Section\,\ref{discuss_bsg}.

Venn et al. \shortcite{venn2000} 
have presented an analysis of three other stars
in M31 (A-F supergiants) and the oxygen abundances for these  
are also plotted in  Figure\,\ref{oxygen_gradient}. Fig.\,\ref{m31image} also 
shows the spatial positions of the A-type supergiants and OB10
in the M31 disk. We see that the
present results strengthen the suggestion of Venn et al.  that the
abundance gradient of oxygen is quite flat, extending the stellar
results into the supposedly metal rich region interior to  R$_{g}=10$\,kpc.
As noted by Venn et al., neglecting the innermost H\,{\sc ii} regions
results in
good agreement between the nebular and stellar results, albeit with
a negligible gradient, so it is
important to try to understand the current discrepancy in the inner
region. 
In the  H\,{\sc ii} region work the observed quantity
used to derive [O/H] is $R_{23}$, the ratio
([OII]$\lambda$3727 + [OIII]$\lambda\lambda$4959, 5007)/H$\beta$,
which is observed to increase with R$_{g}$.  This ratio has
been calibrated as a function of [O/H] by a number of authors,
and indeed as Venn et al. pointed out, using the calibration
of Zaritsky et al. \shortcite{zar94} instead of Pagel et al. 
\shortcite{Pagel79}, reduces the [O/H]
abundance gradient from Blair et al.'s value of $-0.03$ dex\,kpc$^{-1}$ to 
$-0.02$ dex\,kpc$^{-1}$.  Indeed, within the range of $R_{23}$ defined
by the  H\,{\sc ii} regions in question, $0.0 \le R_{23} \le 0.8 $,
there are significant 
differences between the slopes and absolute values of the various
$R_{23}$ versus [O/H] relationships (see Figure 11 of McGaugh 1991). 
In fact McGaugh's own calibration results in an even flatter gradient
than that of Zaritsky et al. \shortcite{zar94}, and systematically 
lower oxygen abundances
by 0.1 to 0.2 dex, with the largest difference being at higher metallicities
(see Kobulnicky et al. 1999, equations 9
and 10 for the parameterization).
For metallicities above solar there are almost
no H\,{\sc ii} regions for which there are independent estimates of
the electron temperature due to the weakness of the [OIII] $\lambda$4363
line. Thus the various calibrations in use depend either upon 
extrapolated empirical relationships, or upon models for which one
must make certain assumptions (for example concerning
abundances, abundance ratios, model atmosphere fluxes and dust).  

Fortunately, Galarza et al. 
\shortcite{gal99} have presented new results for M31 H\,{\sc ii}
regions between 5 -- 15 kpc, and three of these are very close
to the OB association OB\,10 (regions designated K310, K314 and K315). K315
is spatially coincident with OB\,10 and we assume that it is sampling
the nebula excited by its OB stars.  Interestingly, Galarza et al.
\shortcite{gal99}
discuss their H\,{\sc ii} regions by morphological type; center-bright,
ring, complex and diffuse, but only those regions defined as 
center-bright exhibit a significant gradient in $R_{23}$.  Of the
regions mentioned above, only K314 is a center-bright region, it
is about 1.5 arc minutes away from OB\,10-64, and McGaugh's calibration
gives 12+log[O/H] = 8.98, whereas Zaritsky's gives 9.17.  We can apply
these calibrations to all three H\,{\sc ii} regions mentioned above,
plus the 5 sub-apertures centered on K315 for which Galarza et al.
\shortcite{gal99}
also give data, which results in oxygen abundances ranging from 
approximately 8.7 to 9.0 dex using McGaugh's calibration, or
from 8.9 to 9.2 dex using Zaritzsky's (see Table\,\ref{galarza}).  
The former
calibration appears to give 
absolute results in better agreement with the 
stellar results, indeed our absolute abundance of 8.69$\pm$0.1
agrees {\em within the errors} with the McGaugh calibration
of 8.7--9.0\,dex. 
We therefore use this relationship to rederive
oxygen abundances using the published H\,{\sc ii} data from
Blair et al.\,(1982), Dennefeld \& Kunth (1981) and Galarza et al.\
(center-bright regions only). A discussion of which calibration is 
more physically reliable is outside the scope
of this paper, however it seems consistent to 
homogeneously re-derive the abundances from the full observational
dataset using a single relation. We have chosen the McGaugh calibration
simply because it provides better agreement with the OB10 stellar results, 
but would caution against using any one of these calibrations blindly
before more work is carried out on their reliability and the
reasons for discrepancies. 

\begin{figure}
\epsfig{file=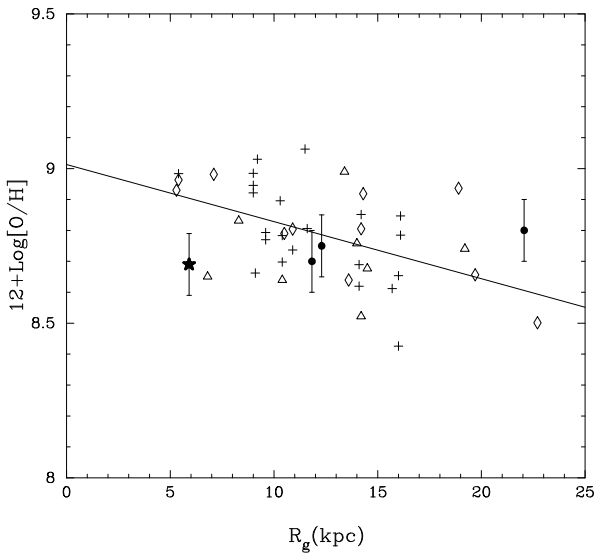,width=8.4cm}
\caption{All H\,{\sc ii} region results from Blair et al.\ (1982; diamonds),
Dennefeld \& Kunth (1981; triangles) and Galarza et al.\ (1999; crosses)
using the calibration of McGaugh (1991). Stellar points are represented
by filled symbols as in Figure 8.  The straight line represents a
least squares fit to the nebular data giving a gradient of 
$-0.018 \pm0.01$ dex\,kpc$^{-1}$. }
\label{allhii}
\end{figure}

These results are shown in
Figure~\ref{allhii} together with the stellar results and the
least squares fit to the nebular results which gives an
oxygen abundance gradient of $-0.018 \pm0.01$ 
dex\,kpc$^{-1}$, with an intercept
at $R_g=0.0$ of 9.02\,dex.   Note that the corresponding
values using Zaritsky's calibration are $-$0.025 dex\,kpc$^{-1}$ and
9.22\,dex,
however the gradient determined from H\,{\sc ii} regions
in the inner, possibly metal rich regions, still depends critically
upon the functional form of the calibration used. Clearly
this problem can only be resolved by more detailed observations
and modelling of these inner regions.  The stellar results
imply that the [O/H] gradient is extremely flat in the
range 5 to 25 kpc.
We have not fitted a gradient
to the four stars as they are not a homogeneous dataset
in terms of their atmospheric parameters and analysis methods;
the star analysed here is a B-type supergiant, those at 12 kpc are
A-type supergiants, while that at 22 kpc is an F-type
supergiant. Certainly, more stellar observations are clearly required, 
as any gradient which does exist may be masked by statistical scatter
or real variations at any particular R$_{g}$. 

\begin{table}
  \caption{{H\,{\sc ii}} regions observed by Galarza et al.\ (1999) 
in the vicinity
of OB\,10; note that K315 is coincident with the position of OB\,10-64 for
which we obtain an oxygen abundance of 8.7\,dex. Derived abundances
[O/H]$_Z$ and [O/H]$_M$ refer to the use of the calibrations
of Zaritsky et al.\ (1994) and McGaugh (1991) respectively.}
\begin{tabular}{l@{\hspace{2mm}}l@{\hspace{2mm}}r@{\hspace{2mm}}r
@{\hspace{2mm}}r}\hline
H\,{\sc ii} region & Morpholgical & log$(R_{23})$ & [O/H]$_Z$ & [O/H]$_M$ \\
identifier & type & & &  \\
\hline
K310  & diffuse & 0.47 & 9.00 & 8.85 \\
K314  & center-bright & 0.24 & 9.17 & 8.98 \\
K315  & complex & 0.39 & 9.07 & 8.90 \\
K315a & diffuse & $<0.23$ & $>9.16$ & $>8.98$ \\
K315b & diffuse & 0.23 & 9.16 & 8.98 \\
K315c & diffuse & 0.16 & 9.20 & 9.01 \\
K315d & ring & 0.49 & 9.00 & 8.83 \\
K315e & ring & 0.63 & 8.84 & 8.69 \\
\hline
\label{galarza}
\end{tabular}
\end{table}

Without the analysis presented here for 
OB\,10-64 we would have assumed that
the initial metallicity of OB\,10-WR1 was much higher than solar, 
and that the wind analysis and stellar evolution discussion would
have been based on quite different parameters. 
This indicates that one must be careful
in assuming an initial metallicity determined from 
global galactic properties when characterising the evolution
of single evolved stars, especially WR stars. It is necessary to 
use more direct methods to determine the state of the natal gas
in the environment of the cluster formation.

\section{Conclusions}

We have carried out detailed, quantitative analyses of 
high-quality spectral data of two massive stars in the young OB\,10
association in M31. This represents the most detailed 
study of hot, massive stars in this galaxy to date, and 
shows that quantitative extragalactic stellar astrophysics
is not only possible with these objects, but is a powerful
diagnostic tool advancing our knowledge in several related
areas. 

Our analyses allows us to probe massive stellar evolution, 
mass loss and stellar wind characteristics in the most massive,
luminous stars in the Local Universe. We show that
we can determine 
accurate chemical abundances in B-type supergiants which 
are probes of the current metallicity of the ISM.
Our stellar analysis indicates that 
the empirical calibration
of the $R_{23}$ ratio in H\,{\sc ii} regions at metallicities
of solar and above is not well constrained. Hence when 
discussing stellar evolution from characteristics of 
evolved massive stars (WR stars, LBV's, blue and red supergiants)
one should be careful to use direct metallicity determinations
of the starforming region -  
photospheric abundance estimates in massive B-type (and A-type, 
see Venn et al. 2000) supergiants provide an excellent 
means of doing so. 

We have demonstrated that it is possible to analyse the physical
and chemical properties of individual WC stars in M31 using 4m class
telescopes, although more precise luminosities and abundances await
UV spectroscopy. Comparison between
OB\,10--WR1 and WCE stars in other galaxies, particularly HD\,92809 in
our Galaxy, reveals that its high luminosity and low C/He ratio
is  more typical of WCE stars in metal {\it poor} environments, in spite of
the low number statistics.
Since OB\,10 is in the inner region of the M31 disk (R$_{g}=5.9$\,kpc), 
we would have wrongly concluded that WR1 originated in a very metal 
{\it rich} region, had we assumed a metallicity from the previous 
H\,{\sc ii} region analysis.

We have carried out a detailed wind-analysis of the B-type supergiant
and (assuming previously determined distances to M31) find its
wind momentum-luminosity relationship to be in excellent 
agreement with similar Galactic early B-type stars. 
This is a further indication (cf Kudritzki et al. 1999), that
a properly calibrated 
wind momentum-luminosity relationship within the Local Group
may allow accurate distance moduli to be determined to 
the most luminous stars in galaxies within $\sim$20\,Mpc. 
In further papers we will analyse a larger set of M31
A and B-type supergiants, extending the radial baseline from 
the central to regions to $R_{\rm g}\ga20$\,kpc. 

\section*{Acknowledgments}
Spectroscopic data were obtained at the William Herschel
Telescope which is operated on the island of La Palma by the 
Isaac Newton Group in the Spanish Observatorio del Roque de los 
Muchachos of the Instituto de Astrofisica de Canarias. 
We are grateful to the staff there for their help in obtaining
the observations, and particularly Nic Walton for assistance 
with some of the Service observing. We acknowledge the 
usefulness and quick response of the ING Service programme 
in getting required extra observations. We thank the 
W.M. Keck foundation and its president Howard B. Keck 
for the generous grant that made the W.M. Keck Observatory
possible. Observations were made through the 
GO programme with the NASA/ESA 
Hubble Space Telescope, and supporting material was 
obtained from the data archive at the Space Telescope Science 
Institute. STScI is operated by the Association of 
Universities for Research in Astronomy, Inc. under 
NASA contract NAS 5-26555.
We gratefully appreciate the use of {\sc cmfgen} courtesy of John Hillier.
Financial support is acknowledged from the Royal Society (PAC),
PPARC (SJS), the Deutsche Zentrum f\"ur Luft-
             und Raumfahrt (DLR) (FB and RPK under grant 50 OR 9909 2), 
the QUB Visiting Fellowship scheme (SJS and DJL)
and the IoA Visiting Fellows scheme (DJL). AH wishes 
 to acknowledge financial support by the spanish DGES
under project PB97-1438-C02-01 and from the Gobierno Autonomo
de Canarias under project PI1999/008.

\bsp

\label{lastpage}

\end{document}